\documentclass[11pt, a4paper]{article}

\usepackage{amsthm}
\usepackage{amsmath}
\usepackage{amsfonts}
\usepackage{booktabs}
\usepackage{dcolumn}
\usepackage{natbib}
\usepackage{hyperref}
\usepackage{fullpage}
\usepackage{graphicx}
\usepackage{natbib}
\usepackage{mathtools}
\usepackage{color}
\usepackage[left = 2.5cm, right = 2.5cm, bottom = 3cm, top = 2.5cm]{geometry}
\usepackage{enumitem}
\usepackage{multirow}
\usepackage{rotating}
\usepackage{calc}
\usepackage{bm}
\usepackage{authblk}
\usepackage{pdfpages}
\usepackage{xr}
\newcommand*{\bb}{\boldsymbol}

\newcommand{\iq}[3]{#1^{#2}_{#3}}

\newtheoremstyle{example}
{3pt} 
{3pt} 
{} 
{0\parindent} 
{\bf}
{:} 
{.5em} 
{} 
\newtheoremstyle{theorem}
{3pt} 
{3pt} 
{\em} 
{0\parindent} 
{\bf}
{:} 
{.5em} 
{} 
\theoremstyle{example} \newtheorem{example}{Example}[section]
\theoremstyle{theorem} 

\def\expect{{\mathop{\rm E}}}
\def\var{{\mathop{\rm var}}}

\def\trace{{\mathop{\rm trace}}}

\title{Empirical bias-reducing adjustments to estimating functions}

\author[1]{Ioannis Kosmidis\thanks{ioannis.kosmidis@warwick.ac.uk}}
\author[2]{Nicola Lunardon}

\affil[1]{Department of Statistics, University of Warwick \authorcr Gibbet Hill Road, Coventry, CV4 7AL, UK}
\affil[2]{Department of Environmental Sciences, Informatics and Statistics \authorcr Ca’ Foscari University of Venice \authorcr Via Torino 150, Venezia Mestre 30170, Italy}

\begin{document}
\maketitle

\begin{abstract}
  We develop a novel and general framework for reduced-bias
  $M$-estimation from asymptotically unbiased estimating
  functions. The framework relies on an empirical approximation of the
  bias by a function of derivatives of estimating function
  contributions. Reduced-bias $M$-estimation operates either
  implicitly, by solving empirically-adjusted estimating equations, or
  explicitly, by subtracting the estimated bias from the original
  $M$-estimates, and applies to models that are partially- or
  fully-specified, with either likelihoods or other surrogate
  objectives. Automatic differentiation can be used to abstract away
  the only algebra required to implement reduced-bias
  $M$-estimation. As a result, the bias reduction methods we introduce
  have markedly broader applicability with more straightforward
  implementation and less algebraic or computational effort than other
  established bias-reduction methods that require resampling or
  evaluation of expectations of products of log-likelihood
  derivatives. If $M$-estimation is by maximizing an objective, then
  there always exists a bias-reducing penalized objective. That
  penalized objective relates closely to information criteria for
  model selection, and can be further enhanced with plug-in penalties
  to deliver reduced-bias $M$-estimates with extra properties, like
  finiteness in models for categorical data.  The reduced-bias
  $M$-estimators have the same asymptotic distribution as the original
  $M$-estimators, and, hence, standard procedures for inference and
  model selection apply unaltered with the improved estimates. We
  demonstrate and assess the properties of reduced-bias $M$-estimation
  in well-used, prominent modelling settings of varying complexity.
\end{abstract}

\section{Introduction}
\subsection{Bias reduction methods}
\label{br_methods}
Reduction of estimation bias in statistical modelling is a task that
has attracted immense research activity since the early days of the
statistical literature. This ongoing activity resulted in an abundance
of general bias-reduction methods with varying levels of
applicability. As is noted in \citet{kosmidis:2014}, the majority of
methods start from an estimator $\hat{\bb \theta}$ of an unknown
parameter ${\bb \theta}$, and attempt to produce an estimator
$\tilde{\bb \theta}$, which approximates the solution of the equation
\begin{equation}
  \label{bias_equation}
  \hat{\bb \theta} - \tilde{\bb \theta} = {\bb B}_G(\bar{\bb \theta}) \, ,
\end{equation}
with respect to $\tilde{\bb \theta}$. In the above, $G$ is the
typically unknown distribution function of the process that generated
the data,
${\bb B}_G({\bb \theta}) = \expect_G(\hat{\bb \theta} - {\bb \theta})$
is the bias function, and $\bar{\bb \theta}$ is the value that
$\hat{\bb \theta}$ is assumed to converge to in probability as
information about ${\bb \theta}$ increases, typically with the volume
of the data. In general, the solution of~\eqref{bias_equation}
requires approximation because both the value of $\bar{\bb \theta}$
and $G$ are unknown or the expectation with respect to $G$ does not
have a closed form.

Table~\ref{br_characteristics} classifies prominent bias-reduction
methods according to various criteria relating to their applicability
and operation. Given the size of the literature on bias-reduction
methods, we only cite key works that shaped or greatly impacted the
area. Bias-reduction methods like vanilla asymptotic bias correction
\citep{efron:1975}, the adjusted scores functions approach in
\citet{firth:1993}, indirect inference in \citet{gourieroux:1993}, and
iterated bootstrap in \citet{kuk:1995} and
\citet{guerrier+dupuis-lozeron+etal:2019}, and the refinements of the
indirect inference principle in \citet{mackinnon+smith:1998} assume
that the model can be fully and correctly specified, in the sense that
$G$ results from the assumed model for specific parameter values. That
assumption allows to either have access to log-likelihood derivatives
and expectations of products of those or to simulate from the
model. In contrast, bias-reduction methods, like jackknife
\citep{quenouille:1956, efron:1982} and bootstrap
\citep{efron+tibshirani:1993, hall+martin:1988} can also apply to at
least partially-specified models; see, also, \citet{newey+smith:2004}
for an asymptotic bias-correction method for partially-specified
models with independent and identically distributed random variables
that applies also to over-identified problems. Such bias reduction
methods can improve estimation in involved modelling settings, where
typically surrogate inference functions are used in an attempt to
either limit the number of hard-to-justify modelling assumptions or
because the full likelihood function is impractical or cumbersome to
compute; see, for example, \citet{wedderburn:1974} for
quasi-likelihood methods, \citet{liang+zeger:1986} for generalized
estimating equations, and \citet{lindsay:1988} and
\citet{varin+reid+firth:2011} for composite likelihood methods.

\citet{kosmidis:2014} classifies bias-reduction methods according to
whether they operate in an explicit or implicit manner when
approximating the solution of~\eqref{bias_equation}. Explicit methods
estimate ${\bb B}_G(\bar{\bb \theta})$ and subtract that estimate from
$\hat{\bb \theta}$. Implicit methods replace
${\bb B}_G(\bar{\bb \theta})$ with
$\hat{{\bb B}}_G(\tilde{\bb \theta})$ for an estimator
$\hat{{\bb B}}_G$ of ${\bb B}_G$, and solve the resulting implicit
equation.

Bias-reduction methods can also be classified according to whether the
necessary approximation of the bias term in~\eqref{bias_equation} is
performed analytically or through simulation. The vanilla
implementations of asymptotic bias correction and adjusted score
functions approximate ${\bb B}_G({\bb \theta})$ with a function
${\bb b}({\bb \theta})$ such that
${\bb B}_G({\bb \theta}) = {\bb b}({\bb \theta}) + O(n^{-3/2})$, where
$n$ is a measure of how the information about ${\bb \theta}$
accumulates.  On the other hand, jackknife, bootstrap, iterated
bootstrap, and indirect inference, generally, approximate the bias by
simulating samples from the assumed model or an estimator of $G$, like
the empirical distribution function. As a result, and depending on how
demanding the computation of $\hat{\bb \theta}$ is, simulation-based
methods are typically more computationally intensive than analytical
methods. Also, implicit, simulation-based methods require special care
and ad-hoc considerations when approximating the solution
of~\eqref{bias_equation}, because the simulation-based estimator of
${\bb B}_G({\bb \theta})$ is not always differentiable with respect to
${\bb \theta}$.

The requirement of differentiation of the estimating or objective 
functions for some of the bias-reduction methods in
Table~\ref{br_characteristics} has also resulted in considerable
analytical effort over the years (see, for example,
\citealt{kosmidis+firth:2009} for multivariate generalized nonlinear
models, and \citealt{grun+kosmidis+zeileis:2012} for Beta regression
models). Nevertheless, differentiation is nowadays a task requiring
increasingly less analytical effort because of the availability of
comprehensive automatic differentiation routines
\citep{griewank+walther:2008} in popular computing environments. Such
routines can be found, for example, in the \texttt{ForwardDiff} Julia
package \citep{revels+lubin+papamarkou:2016}, and the C++ package
\texttt{CppAD} \citep{bell:2021} that enabled the development of a
range of template modelling software, like the \texttt{TMB} R package
\citep{kristensen+nielsen+berg+etal:2016, rproject}. Even if the
differentiation effort can be partly mitigated, the vanilla versions of
asymptotic bias correction in \citet{efron:1975} and the bias-reducing
adjusted score functions in \citet{firth:1993} require the computation
of expectations of products of log-likelihood derivatives under the
model. Those expectations are intractable or expensive to compute for
models with intractable or cumbersome likelihoods, and can be hard to
derive even for relatively simple models; see, for example,
\citet[Section~2.3]{grun+kosmidis+zeileis:2012} for how involved those
expectations are for Beta regression models.

\begin{table}[t]
  \caption{Classification of bias-reduction methods of general
    applicability (Method). The classification is based on the
    level of model specification (Model), the way the method
    approximates the bias (${\bb B}_G(\bar{\bb \theta})$), the type
    (Type) of the method according to the classification in
    \citet{kosmidis:2014}, and on the method's requirements in terms
    of computation of expectations ($\expect(\cdot)$), differentiation
    ($\partial \cdot$), and access to the original estimator
    ($\hat{\bb \theta}$). Key references include:
    \citet[Section~10]{efron:1975} and \citet{cordeiro+mccullagh:1991}
    for asymptotic bias correction; \citet{firth:1993} and
    \citet{kosmidis+firth:2009} for adjusted score functions;
    \citet{efron+tibshirani:1993} and \citet{hall+martin:1988} for
    bootstrap; \citet{quenouille:1956} and \citet{efron:1982} for
    jackknife; \citet{gourieroux:1993} for indirect inference,
    \citet{kuk:1995} and \citet{guerrier+dupuis-lozeron+etal:2019} for
    iterated bootstrap, and \citet{mackinnon+smith:1998} for
    refinements of the indirect inference principle; and this manuscript for RB$M$-estimation.}
  \begin{center}
  \begin{tabular}{lccccccc}
    \toprule
    & & & & \multicolumn{3}{c}{Requirements} \\ \cmidrule{5-7}
    Method & Model & ${\bb B}_G(\bar{\bb \theta})$ & Type & $\expect(\cdot)$ & $\partial \cdot$ & $\hat{\bb \theta}$  \\ \midrule
    Asymptotic bias correction & full & analytical & explicit & yes & yes & yes \\ 
    Adjusted score functions & full & analytical & implicit & yes & yes  & no \\
    Bootstrap & partial & simulation & explicit & no  & no & yes \\
    Jackknife & partial & simulation & explicit & no & no  & yes \\ %
    Iterated bootstrap / Indirect inference & full & simulation & implicit & no & no  & yes \\ \midrule
    Explicit RB$M$-estimation & partial & analytical & explicit & no & yes & yes \\
    Implicit RB$M$-estimation & partial & analytical & implicit & no & yes & no \\
    \bottomrule
  \end{tabular}
\end{center}
\label{br_characteristics}
\end{table}

Finally, except of the adjusted scores approach in \citet{firth:1993},
all the bias-reduction methods reviewed in
Table~\ref{br_characteristics} require the original estimator
$\hat{\bb \theta}$ and they cannot operate without it. For this
reason, they directly inherit any of the instabilities that
$\hat{\bb \theta}$ may have. For example, in multinomial logistic
regression, there is always a positive probability of data separation
\citep{albert+anderson:1984} that results in infinite maximum
likelihood estimates. Then, asymptotic bias correction, bootstrap,
iterated bootstrap, and jackknife cannot be applied. The direct
dependence on $\hat{\bb \theta}$ may be more consequential for naive
implementations of the latter three methods because they are
simulation-based; even if data separation did not occur for the
original sample, there is always positive probability that it occurs
for at least one of the sub-samples or simulated samples. There is no
easy way of knowing this before carrying out the simulation, and
boundary estimates, if they can be identified reliably, can only be
handled in an ad-hoc way.

\subsection{Reduced-bias $M$-estimation}

The current work develops a novel approach to the reduction of the
asymptotic bias of $M$-estimators from approximately unbiased
estimating functions. We call the new estimators reduced-bias
$M$-estimators, or RB$M$-estimators in short. As noted in the last two
rows of Table~\ref{br_characteristics}, RB$M$-estimation applies to
models that are at least partially-specified and does not require the
computation of any expectations. The analytical approximation to the
bias function required by RB$M$-estimation relies only on derivatives
of the contributions to the estimating functions. RB$M$-estimators
with bias of lower asymptotic order than the original $M$-estimators
result either in an explicit or implicit manner. Explicit
RB$M$-estimation proceeds by directly subtracting the value of the
analytical approximation to the bias function at the $M$-estimates
from the $M$-estimates. Implicit RB$M$-estimation, on the other hand,
does not require the original $M$-estimator, and instead relies on an
additive, bounded-in-probability adjustment to the estimating
functions. The key difference to the established implicit
\citep{firth:1993, gourieroux:1993, kuk:1995} and explicit
bias-reduction methods (bootstrap, jackknife, asymptotic bias
correction), is that the empirical adjustments introduced here depend
only on the first two derivatives of the contributions to the
estimating functions. Hence, they require neither the computation of
cumbersome expectations nor the, potentially computation-intensive and
error-prone calculation of $M$-estimates from simulated samples.

The RB$M$-estimators have the same asymptotic distribution, and hence
the same asymptotic efficiency properties, as the original
$M$-estimators. The definition of asymptotically-valid inference and
model selection procedures follows directly from the ones developed
for the original $M$-estimators (e.g. generalized Wald tests and
information criteria).

Implicit and explicit RB$M$-estimation, apart from being more general
than the methods listed in Table 1, are also easier to implement for
arbitrary models through automatic differentiation. The only key
required input for a computer implementation is an implementation of
the contributions to either the estimating or objective function. The
\texttt{MEstimation} Julia package \citep{mestimation} is a
proof-of-concept of such a general-purpose implementation.

If the estimating functions are the components of the gradient of an
objective function, as is the case in estimation based on likelihoods
or composite likelihoods, then, we show that implicit RB$M$-estimation
can always be achieved by the maximization of an appropriately
penalized version of the objective. This is in contrast to the method
in \citet{firth:1993} which does not always have a penalized
likelihood interpretation; see, for example,
\citet[Theorem~1]{kosmidis+firth:2009}. Moreover, it is shown that the bias-reducing penalized objective
closely relates to information criteria for model selection based on
the Kullback-Leibler divergence. The penalties that are used for bias
reduction and model selection differ only by a known scalar
constant. These observations, establish, for the first time, a strong
link between model selection and reduction of bias in estimation. We
also show how plug-in penalties to the penalized objectives can be
used to enrich RB$M$-estimation with extra, desirable properties, such
as estimates that are always finite in categorical response models,
without sacrificing reduction of bias.

Section~\ref{modelling_setting} introduces notation, the general
modelling setting we consider, and the assumptions underpinning the
theoretical developments. In Section~\ref{sec:theory}, we derive the
leading term of the bias of the $M$-estimator, give the bias-reducing
adjustments to estimating functions and their empirical versions, and
describe the explicit and implicit versions of RB$M$-estimation.
Section~\ref{sec:inference} shows that the asymptotic distribution of
the RB$M$-estimator is the same as that of the original $M$-estimator,
and introduces Wald-type and generalized score approximate pivots that
can be used for inference. Section~\ref{sec:penalties} shows that
implicit RB$M$-estimation can always be achieved using empirical
bias-reducing penalties to objective functions, draws links with model
selection, and discusses the enrichment of the penalized objectives
with plug-in penalties that endow RB$M$-estimators with extra
properties. Section~\ref{sec:highdim} explores the effectiveness of
RB$M$-estimation in partially-specified, high-dimensional regression
settings, and Section~\ref{spatiotemporal} discusses RB$M$-estimation
in models with independent random variables to models for temporally
or spatially correlated observations.
We demonstrate and assess the breath and properties of the
RB$M$-estimation framework with examples from well-used, important
modelling settings of increasing complexity, including: estimation the
ratio of two means with minimal distributional assumptions
(Example~\ref{ratio_estimation} and Example~\ref{ratio_estimation2}),
estimation and model selection in generalized linear models
(GLMs; Examples~\ref{glm_example}, \ref{probit_bias},
and~\ref{probit_model_selection}), composite likelihood methods for
the estimation of Gaussian max-stable processes
(Example~\ref{max_stable_ex}), estimation of autoregressive models of
order one (Example~\ref{OLS_AR}), and estimation and uncertainty
quantification for autologistic regression models for the analysis of
spatially clustered data (Examples~\ref{autologistic}
and~\ref{highdim_autologistic}). See, also
Section~\ref{sec:negbin_bias} in the Supplementary Material document for
an empirical assessment of RB$M$-estimators in negative binomial
regression with many covariates. Section~\ref{sec:discussion}
concludes with discussion on the developments and possible extensions
of this work.

\section{Modelling setting and assumptions}
\label{modelling_setting}

\subsection{Estimating functions}
\label{EF}

Suppose we observe realizations ${\bb y}_1, \ldots, {\bb y}_k$ of
a sequence of random vectors ${\bb Y}_1, \ldots, {\bb Y}_k$ with
${\bb y}_i = (y_{i1}, \ldots, y_{ic_i})^\top \in \mathcal{Y} \subset
\Re^{c_i}$, possibly with a sequence of covariate vectors
${\bb x}_1, \ldots, {\bb x}_k$, with
${\bb x}_i = (x_{i1}, \ldots, x_{iq_i})^\top \in \mathcal{X} \subset
\Re^{q_i}$, and $c_i \ge 1$ and $q_i \ge 1$. We assume that
${\bb Y}_1, \ldots, {\bb Y}_k$ are independent conditionally on
${\bb x}_1, \ldots, {\bb x}_k$.  Let
${\bb Y} = ({\bb Y}_1^\top, \ldots, {\bb Y}_k^\top)^\top$, and $X$ be
the set of ${\bb x}_1, \ldots, {\bb x}_k$, and denote by
$G \equiv G({\bb Y} \mid X)$ the typically unknown, underlying joint
distribution function.

One of the typical aims in statistical modelling is to estimate at
least a sub-vector of an unknown $p$-dimensional parameter
${\bb \theta} \in {\bb \Theta} \subset \Re^p$ using data
${\bb y}_1, \ldots, {\bb y}_k$ and ${\bb x}_1, \ldots, {\bb
  x}_k$. This is most commonly achieved through an $M$-estimator
$\hat{\bb \theta}$ \citep[][Chapter~5]{vaart:1998} that results from
the solution of a system of $p$ estimating equations
\begin{equation}
  \label{estimating_equations}
  \sum_{i = 1}^k{\bb \psi}^i({\bb \theta}) = {\bb 0}_p \, ,
\end{equation}
with respect to ${\bb \theta}$. In
equation~\eqref{estimating_equations}, ${\bb 0}_p$ is a $p$-vector of
zeros,
${\bb \psi}^i({\bb \theta}) = (\psi_1^i({\bb \theta}), \ldots,
\psi_{p}^i({\bb \theta}))^\top$,
${\bb \psi}^i({\bb \theta}) = {\bb \psi}({\bb \theta}, {\bb Y}_i, {\bb
  x}_i)$, and
$\psi^i_r({\bb \theta}) = \psi_r({\bb \theta}, {\bb Y}_i, {\bb x}_i)$,
$r \in \{1, \ldots, p\}$.  Examples of estimation methods that fall
within the above framework are estimation via quasi-likelihood methods
\citep{wedderburn:1974} and generalized estimating equations
\citep{liang+zeger:1986}. \citet{stefanski+boos:2002} provide an
accessible overview of estimating functions and demonstrate their
generality and the handling of challenging estimation problems with
less assumptions than likelihood-based estimation.

One way to derive estimating functions is through a, typically
stronger, modelling assumption that ${\bb Y}_i$ has a distribution
function with joint mixed density
$f_i({\bb y}_i \mid {\bb x}_i, {\bb \theta})$ or that the joint mixed
densities of sub-vectors of ${\bb Y}_i$ can be composed into a
composite function $f_i({\bb y}_i \mid {\bb x}_i, {\bb \theta})$. The
estimator $\hat{\bb \theta}$ can then be taken to be the maximizer of
the objective function
\begin{equation}
  \label{objective_function}
  l({\bb \theta}) = \sum_{i =1}^k \log f_i({\bb y}_i \mid {\bb x}_i, {\bb \theta}) \, .
\end{equation}
The word ``mixed'' is used here to allow some of the components of
${\bb y}_i$ to be continuous and some discrete. If the objective
function~\eqref{objective_function} is used, then the estimating
functions in~\eqref{estimating_equations} have
${\bb \psi}^i({\bb \theta}) = \nabla \log f_i({\bb y}_i \mid {\bb
  x}_i, {\bb \theta})$, assuming that the gradient exists in
$\bb \Theta$. Prominent examples of estimation methods that involve an
objective function of the form~\eqref{objective_function} are maximum
likelihood and maximum composite likelihood \citep[see, for
example,][]{lindsay:1988, varin+reid+firth:2011}.

\subsection{Assumptions}
\label{assumptions}
The sufficient conditions we employ for the theoretical developments
in this work are standard and cover a wealth of $M$-estimation
problems. The assumptions are listed in detail in
Section~\ref{sec:assumptions_sup} of the Supplementary Material document
and include: the unbiasedness of the estimating function and
convergence of the $M$-estimator to the root of the expectation of the
estimating function with respect to $G$
(assumption~\ref{consistency}), local smoothness of the estimating
functions and existence of a sufficient number of derivatives of those
around that root (assumption~\ref{smoothness}), existence of moments
of products of derivatives of estimating functions under the unknown
data generating process along with conditions on their growth rate in
terms of the amount of information about the parameter
(assumptions~\ref{derivative_orders}-\ref{invertibility}). Many of
those conditions can be derived from others for particular models and
estimation methods. An annotated analysis of the links among them, and
their wide scope and applicability is provided in
Section~\ref{sec:assumptions_sup} of the Supplementary Material
document. Unless otherwise stated, in what follows, we assume that
these assumptions hold.

\section{Adjusted estimating equations for bias reduction}
\label{sec:theory}

\subsection{Asymptotic bias}
\label{asymptotic_bias}

It can be shown that the bias of the $M$-estimator $\hat{\bb \theta}$
is $\expect_G(\hat{\bb \theta} - \bar{\bb \theta}) = O(n^{-1})$, where
$\bar{\bb \theta}$ is such that
$\expect_G ({\bb \psi}^i(\bar{\bb \theta})) = {\bb 0}_p$ for all
$i \in \{1, \ldots, k\}$.  This provides some reassurance that, as the
information about the parameter ${\bb \theta}$ grows, estimation bias
vanishes. Nevertheless, the finite sample bias of $\hat{\bb \theta}$
is typically not zero.  It is also possible to write down the bias in
the more specific form
$\expect_G(\hat{\bb \theta} - \bar{\bb \theta}) = {\bb b}(\bar{\bb
  \theta}) + O(n^{-3/2})$, where ${\bb b}(\bar{\bb \theta})$ depends
on joint moments, with respect to $G$, of estimating functions and
derivatives of those. Then, because
${\bb b}(\hat{\bb \theta}) \stackrel{p}{\longrightarrow} {\bb
  b}(\bar{\bb \theta})$ when
$\hat{\bb \theta} \stackrel{p}{\longrightarrow} \bar{\bb \theta}$, we
can define a reduced-bias estimator
$\hat{\bb \theta} - {\bb b}(\hat{\bb \theta})$.

The estimator $\hat{\bb \theta} - {\bb b}(\hat{\bb \theta})$ has been
shown to, indeed, have better bias properties when estimation is by
maximum likelihood (ML) and the model is correctly specified
\citep[see][Section~10]{efron:1975}. When the model is correctly
specified, the unknown joint distribution function $G({\bb Y} \mid X)$
is assumed to be a particular member of the family of distributions
specified fully by $f_i({\bb y}_i \mid {\bb x}_i, {\bb \theta})$ when
forming~\eqref{objective_function}. It is, then, possible to evaluate
${\bb b}({\bb \theta})$ and, hence, compute
$\hat{\bb \theta} - {\bb b}(\hat{\bb \theta})$ in light of data,
because the expectations involved in the joint null moments are with
respect to the modelling assumption. This is the basis of more refined
bias reduction methods, like the adjusted score function approach that
has been derived in \citet{firth:1993} and explored further in
\citet{kosmidis+firth:2009}.

In the more general setting of Section~\ref{modelling_setting}, where
the model can be only partially specified, naive evaluation of
${\bb b}({\bb \theta})$ using
$f_i({\bb y}_i \mid {\bb x}_i, {\bb \theta})$ not only does not lead
to reduction of bias, in general, but it can also inflate the bias;
see, for example, \citet[Section~2.1]{lunardon+scharfstein:2017}, and
Example~\ref{OLS_AR} for the estimation of autoregressive processes.
Even if the researcher is comfortable to assume that
$f_i({\bb y}_i \mid {\bb x}_i, {\bb \theta})$ is correctly specified,
the applicability of standard bias-reduction methods is hampered
whenever $f_i({\bb y}_i \mid {\bb x}_i, {\bb \theta})$ is impossible
or impractical to compute in closed form. Estimation and inference, in
those cases, is typically based on pseudo likelihoods or estimating
functions. An example of this kind is within the framework of
max-stable processes for which the evaluation of the joint density
becomes quickly infeasible when the number of site locations increases
\citep{davison+gholamrezaee:2012}; see, also,
Example~\ref{max_stable_ex}.

\subsection{Family of bias-reducing adjustments to estimating functions}

Consider the estimator $\tilde{\bb \theta}$ that results from the
solution of the adjusted estimating equations
\begin{equation}
  \label{adjusted_estimating_equations}
  \sum_{i=1}^k {\bb \psi}^i({\bb \theta}) + {\bb A}({\bb \theta}) = {\bb 0}_p \, ,
\end{equation}
where both ${\bb A}({\bb \theta}) = {\bb A}({\bb \theta}, {\bb Y}, X)$
and its derivatives with respect to ${\bb \theta}$ are $O_p(1)$ as $n$
grows.
A calculation similar to that in \citet[Section~7.3]{mccullagh:2018}
leads to a stochastic Taylor expansion for
$\tilde{\bb \theta} - \bar{\bb \theta}$ (see expression~(\ref{taylor})
in Section~\ref{sec:supp_taylor} of the Supplementary Material
document), whose expectation with respect to $G$ gives that the bias
of $\tilde{{\bb \theta}}$ is
\begin{equation}
  \label{explicit_bias}
  \expect_G{(\tilde{\theta}^{r} - \bar{\theta}^{r})} = -\iq{\mu}{ra}{}
  \expect_G (A_a) + \frac{1}{2} \iq{\mu}{ra}{} \iq{\mu}{bc}{}
  \left(2 \iq{\nu}{}{ab,c} - \iq{\mu}{de}{} \iq{\nu}{}{c,e}
    \iq{\mu}{}{abd} \right) + O(n^{-3/2})\, .
\end{equation}
In the above expression, we employ index notation. The quantities
$\iq{\mu}{}{R_a}({\bb \theta}) = \expect_G(l_{R_a}({\bb \theta}))$
with
$l_{R_a}({\bb \theta}) = \sum_{i=1}^k \partial^{a-1} \psi_{r_1}^i({\bb
  \theta}) / \partial \theta^{r_2} \cdots \partial \theta^{r_a}$ are
expectations of derivatives of the estimating function,
$(R_a = \{r_1, \ldots, r_a\}; r_j \in \{1, \ldots, p\}$),
$\iq{\mu}{ra}{}({\bb \theta})$ is the inverse of
$\iq{\mu}{}{ar}({\bb \theta})$, and
$\iq{\nu}{}{c,e}({\bb \theta}) = \expect_G(l_{c}({\bb
  \theta})l_{e}({\bb \theta}))$ and
$\iq{\nu}{}{ab,c}({\bb \theta}) = \expect_G(l_{ab}({\bb
  \theta})l_{c}({\bb \theta}))$. Unless otherwise stated, the
dependence on the parameter is omitted whenever quantities are
evaluated at $\bar{\bb \theta}$, as is the case in the right-hand side
of~(\ref{explicit_bias}).  For example,
$\iq{\mu}{ra}{} = \iq{\mu}{ra}{}(\bar{\bb
  \theta})$. Expansion~\eqref{explicit_bias} implies that an estimator
$\tilde{\bb \theta}$ with bias
$\expect_G{(\tilde{\theta}^{r} - \bar{\theta}^{r})} = O(n^{-3/2})$ is
obtained by the solution of~\eqref{adjusted_estimating_equations} with
any adjustment ${\bb A}(\bb \theta)$ satisfying
\begin{equation}
  \label{adjustment}
  \expect_G(A_r(\bb \theta)) = \frac{1}{2} \iq{\mu}{ab}{}(\bb \theta) \left\{ 2 \iq{\nu}{}{ra,b}(\bb \theta) - \iq{\mu}{cd}{}(\bb \theta) \iq{\nu}{}{b,d}(\bb \theta) \iq{\mu}{}{rac}(\bb \theta) \right\} + O(n^{-1/2}) \, .
\end{equation}
Then $\tilde{\bb \theta}$ has, asymptotically, smaller bias than
$\hat{\bb \theta}$.
In other words, expression~\eqref{adjustment} defines a family of
bias-reducing adjustments to estimating functions.
Expansion~\eqref{explicit_bias}, and, hence, all results below remain
valid if the estimating functions are asymptotically unbiased
with $O(n^{-1/2})$ expectation.

An obvious candidate for ${\bb A}({\bb \theta})$ has $r$th component
the first term in the right-hand side of~\eqref{adjustment}. If
$f_i({\bb y}_i \mid {\bb x}_i, {\bb \theta})$ fully specifies $G$ and
the estimation method is ML, then the Bartlett relation
$\iq{\mu}{}{cd}({\bb \theta}) + \iq{\nu}{}{c, d}({\bb \theta}) = 0$
holds \citep[see, for example,][Section~9.2]{pace+salvan:1997}, and
the adjustment satisfying~\eqref{adjustment} becomes
$\iq{\mu}{ab}{}(\bb \theta) \{ 2 \iq{\nu}{}{ra,b}(\bb \theta) +
\iq{\mu}{}{rab}(\bb \theta) \}/2$. The latter expression is the
bias-reducing adjustment for ML derived in \citet{firth:1993} for
fully-specified models.

For the more general setting of Section~\ref{modelling_setting},
however, the underlying distribution $G$ is typically only partially
specified through $f_i({\bb y}_i \mid {\bb x}_i, {\bb \theta})$ or
relations of moments of the random component. Hence, the expectations
involved in the right-hand side of~\eqref{adjustment} cannot be
computed.

\subsection{Empirical bias-reducing adjustments to estimating functions}
\label{empirical_adjustments}

The weak law of large numbers and a straightforward calculation can be
used to produce an easy-to-implement family of empirical adjustments
that delivers bias reduction of $M$-estimators in the general
framework of Section~\ref{modelling_setting}.
In particular, the empirical adjustment of the form
\begin{equation}
  \label{general_empirical_adjustment}
  A_r({\bb \theta}) = \frac{1}{2} \iq{l}{ab}{}({\bb \theta}) \left\{ 2 \iq{l}{}{ra|b}({\bb \theta}) - \iq{l}{cd}{}({\bb \theta}) \iq{l}{}{b|d}({\bb \theta}) \iq{l}{}{rac}({\bb \theta})\right\} \, ,
\end{equation}
satisfies equation~\eqref{adjustment}. In the above expression,
$l_{r|s}(\bb \theta) = \sum_{i = 1}^k \psi_r^i (\bb \theta) \psi_s^i
(\bb \theta)$ and
$l_{rs|t}(\bb \theta) = \sum_{i = 1}^k \{ \partial \psi_r^i({\bb
  \theta}) / \partial \theta^s \}\psi_t^i({\bb \theta})$ are
estimators of
$\nu_{r, s}(\bb \theta) = \expect_G(\sum_{i = 1}^k \psi^i_r(\bb
\theta) \sum_{j = 1}^k\psi^j_s(\bb \theta))$ and
$\nu_{rs, t}(\bb \theta) = \expect_G(\sum_{i = 1}^k (\partial
\psi^i_r(\bb \theta) / \partial \theta^s) \sum_{j = 1}^k \psi^j_t(\bb
\theta))$, respectively, $\iq{l}{st}{}({\bb \theta})$ is the matrix
inverse of $\iq{l}{}{ts}({\bb \theta})$ and an estimator of
$\iq{\mu}{rs}{}({\bb \theta})$, and $\iq{l}{}{stu}({\bb \theta})$ is
an estimator of $\iq{\mu}{}{stu}({\bb \theta})$. The matrix form of
expression~\eqref{general_empirical_adjustment} sets the $r$th element
of the vector of empirical bias-reducing adjustments to
\begin{equation}
  \label{general_empirical_adjustment_matrix}
  A_r({\bb \theta}) = - \trace\left\{{\bb j}({\bb \theta})^{-1} {\bb d}_r({\bb \theta}) \right\} -
  \frac{1}{2}\trace\left[ {\bb j}({\bb \theta})^{-1} {\bb e}({\bb \theta}) \left\{{\bb j}({\bb \theta})^{-1}\right\}^\top {\bb u}_r({\bb \theta}) \right] \, ,
\end{equation}
where
${\bb u}_r({\bb \theta})=\sum_{i=1}^k \nabla \nabla^\top \psi_r^i({\bb
  \theta})$, and ${\bb j}({\bb \theta})$ is the matrix with $s$th row
$-\sum_{i=1}^k \nabla \psi_s^i({\bb \theta})$. The matrix
${\bb j}({\bb \theta})$, assumed to be invertible but is not
necessarily symmetric, coincides with the negative of the hessian
matrix $\nabla \nabla^\top l({\bb \theta})$ when estimation is through
the maximization of an objective function like
\eqref{objective_function}, and is the observed information in maximum
likelihood estimation. The matrices ${\bb e}({\bb \theta})$ and
${\bb d}_r({\bb \theta})$ correspond to the quantities
$\iq{l}{}{r|s}({\bb \theta})$ and $\iq{l}{}{rs|t}({\bb \theta})$
in~\eqref{general_empirical_adjustment}, respectively.

\subsection{Explicit and implicit reduced-bias $M$-estimation}

According to the classification of bias-reduction methods in
\citet{kosmidis:2014}, the solution of the adjusted estimating
equations
$\sum_{i=1}^k{\bb \psi}^i({\bb \theta}) + {\bb A}({\bb \theta}) = {\bb
  0}_p$, with ${\bb A}({\bb \theta})$ as
in~\eqref{general_empirical_adjustment_matrix}, results in an implicit
bias-reduction method. Hence, the resulting estimator
$\tilde{\bb \theta}$ is called the implicit
RB$M$-estimator. From expression~\eqref{explicit_bias},
another estimator with $o(n^{-1})$ bias is
\begin{equation}
  \label{bias_correction}
  {\bb \theta}^\dagger = \hat{\bb \theta} + {\bb j}(\hat{\bb \theta})^{-1} {\bb A}(\hat{\bb \theta}) \, .
\end{equation}
The estimator~\eqref{bias_correction} defines an explicit
bias-reduction method for partially-specified models, and is called
the explicit RB$M$-estimator. Both explicit and implicit
RB$M$-estimation are attractive compared to other bias-reduction
methods whose applicability is limited to either cases where
$\sum_{i=1}^k{\bb \psi}^i({\bb \theta})$ is the gradient of the
log-likelihood function of a correctly-specified model \citep[see,
e.g.][]{cordeiro+mccullagh:1991, firth:1993}, or samples from a
correctly-specified model can be simulated \citep[see,
e.g.][]{gourieroux:1993, kuk:1995,
  guerrier+dupuis-lozeron+etal:2019}. Also, this generalization comes
with less implementation requirements because both explicit and
implicit RB$M$-estimation require only the estimating functions and
the first two derivatives of those.  The RB$M$-estimators can directly
be computed and used in settings that involve realizations of $k$
independent random vectors with dependent components, like the
generalized estimating equations in \citet{liang+zeger:1986} and
composite likelihood methods \citep{varin+reid+firth:2011}.

The choice between using ${\bb \theta}^\dagger$ and
$\tilde{\bb \theta}$ is application-dependent. For example,
$\tilde{\bb \theta}$ in Example~\ref{ratio_estimation} below has
certain robustness properties that ${\bb \theta}^\dagger$
lacks. Furthermore, as is shown in Section~\ref{sec:penalties}, in
$M$-estimation problems that involve an objective function
$\tilde{\bb \theta}$ can be viewed as the maximizer of a bias-reducing
penalized objective that has strong connections to model selection
procedures. On the other hand, for general models for which
${\bb \theta}^\dagger$ and $\tilde{\bb \theta}$ are not available in
closed form, ${\bb \theta}^\dagger$ is typically faster to compute
because it results from a single step of a quasi Newton-Raphson
iteration with stationary point $\tilde{\bb \theta}$ (see
Section~\ref{sec:implementation} in the Supplementary Material
document).  The \texttt{MEstimation} Julia package implements explicit
and implicit RB$M$-estimation for arbitrary models using automatic
differentiation.

\begin{example}
  \label{ratio_estimation}
    
  {\bf Ratio of two means} Consider a setting where independent random
  pairs $(X_1, Y_1), \ldots, (X_n, Y_n)$ are observed, and suppose
  that interest is in the ratio of the mean of $Y_i$ to the mean of
  $X_i$, that is $\theta = \mu_Y / \mu_X$, with
  $\mu_Y = \expect_G(Y_i)$ and $\mu_X = \expect_G(X_i) \ne 0$
  $(i = 1, \ldots, n)$.

  Assuming that sampling is from an infinite population, one way of
  estimating $\theta$ without any further assumptions about the joint
  distribution of $(X_i, Y_i)$ is to set up an unbiased estimating
  equation of the form~\eqref{estimating_equations}, with
  $\psi^i(\theta) = Y_i - \theta X_i$. Then, the $M$-estimator is
  \begin{equation}
    \label{M_ratio}
    \hat\theta = \arg \mathop{\rm solve}_{\theta \in \Re} \left\{\sum_{i = 1}^n \psi^i(\theta) = 0 \right\} = \frac{s_Y}{s_X} \, ,
  \end{equation}
  where $s_X = \sum_{i = 1}^n X_i$ and $s_Y = \sum_{i = 1}^n Y_i$. The
  estimator $\hat\theta$ is generally biased and efforts have been
  made in reducing its bias; see, for example, \citet{durbin:1959},
  for an early work in that direction using the jackknife.
 
  For the estimating function in~\eqref{M_ratio}, $j(\theta) = s_X$,
  $u(\theta) = 0$,
  $e(\theta) = s_{YY} + \theta^2 s_{XX} - 2\theta s_{XY}$, and
  $d(\theta) = -s_{XY} + \theta s_{XX}$, where
  $s_{XX} = \sum_{i = 1}^n X_i^2$, $s_{YY} = \sum_{i = 1}^n Y_i^2$ and
  $s_{XY} = \sum_{i = 1}^n X_i Y_i$. So, the empirical bias-reducing
  adjustment in~\eqref{general_empirical_adjustment} is
  $s_{XY}/s_{X} - \theta s_{XX}/s_{X}$. The solution of the adjusted
  estimating equation $\sum_{i = 1}^n \psi^i(\theta) + A(\theta) = 0$
  results in the implicit RB$M$-estimator
  $\tilde\theta = (s_Y + s_{XY}/s_{X})/(s_X + s_{XX}/s_{X})$.
  A direct substitution of $j(\theta)$ and $A(\theta)$
  in~\eqref{bias_correction} gives that the explicit RB$M$-estimator
  has the form  $\theta^\dagger = \hat\theta (1 - s_{XX}/s_{X}^2) + s_{XY}/s_X^2$.

  \begin{table}[t!]
    \caption{Simulation-based estimates of the bias (B), mean squared
      error (MSE), mean absolute error (MAE), and probability of
      underestimation (PU) for the estimation of a ratio of means in
      the setting of Example~\ref{ratio_estimation}. The estimator
      $\theta_n$ is either the $M$-estimator $\hat\theta$, the
      Jackknife estimator $\theta^{(J)}$, the ``better''
      nonparametric bootstrap estimators $\theta^{(B)}$ in
      \citet[Section~10.4]{efron+tibshirani:1993} using $500$
      resamples, the explicit RB$M$-estimator $\theta^\dagger$, or the
      implicit RB$M$-estimator $\tilde\theta$. Figures are reported in
      2 decimal places, and a figure of $0.00$ indicates an estimated
      bias between $-0.005$ and $0.005$. The simulation error for the
      bias estimates is in $(0.001, 0.002)$.}
    \begin{center}
      \begin{tabular}{ll
        D{.}{.}{2}D{.}{.}{2}D{.}{.}{2}D{.}{.}{2}D{.}{.}{2}D{.}{.}{2}D{.}{.}{2}}
        \toprule
        & \multirow{2}{*}{$\theta_n$} & \multicolumn{6}{c}{$n$} \\
        \cmidrule{3-9}
        \multicolumn{1}{c}{} & & \multicolumn{1}{c}{5} & \multicolumn{1}{c}{10} & \multicolumn{1}{c}{20} & \multicolumn{1}{c}{40} & \multicolumn{1}{c}{80} & \multicolumn{1}{c}{160} & \multicolumn{1}{c}{320} \\
        \midrule
 \multirow{5}{*}{B} &                $\hat\theta$ &  1.19 &  0.53 &  0.25 &  0.12 & 0.06 & 0.03 &  0.01 \\
 &         $\theta^{(J)}$ & -0.38 & -0.07 & -0.01 & 0.00 & 0.00 & 0.00 & 0.00 \\
 & $\theta^{(B)}$ & -0.19 &  0.01 &  0.01 &  0.00 & 0.00 & 0.00 & 0.00 \\
 &                $\theta^\dagger$ &  0.62 &  0.17 &  0.05 &  0.01 & 0.00 & 0.00 &  0.00 \\
 &                $\tilde\theta$ &  0.40 &  0.10 &  0.03 &  0.01 & 0.00 & 0.00 & 0.00 \\ \midrule
 \multirow{5}{*}{MSE} &                $\hat\theta$ & 12.99 &  3.76 &  1.47 &  0.65 & 0.30 & 0.15 &  0.07 \\
 &         $\theta^{(J)}$ & 12.69 &  3.04 &  1.29 &  0.61 & 0.29 & 0.15 &  0.07 \\
 & $\theta^{(B)}$ & 59.22 &  3.00 &  1.29 &  0.61 & 0.29 & 0.15 &  0.07 \\
 &                $\theta^\dagger$ & 10.09 &  3.11 &  1.31 &  0.61 & 0.29 & 0.15 &  0.07 \\
 &                $\tilde\theta$ &  9.37 &  3.04 &  1.30 &  0.61 & 0.29 & 0.15 &  0.07 \\ \midrule
 \multirow{5}{*}{MAE} &                $\hat\theta$ &  2.23 &  1.37 &  0.91 &  0.62 & 0.43 & 0.30 &  0.22 \\
 &         $\theta^{(J)}$ &  2.29 &  1.31 &  0.88 &  0.61 & 0.43 & 0.30 &  0.21 \\
 & $\theta^{(B)}$ &  2.20 &  1.29 &  0.88 &  0.61 & 0.43 & 0.30 &  0.21 \\
 &                $\theta^\dagger$ &  2.03 &  1.29 &  0.88 &  0.61 & 0.43 & 0.30 &  0.21 \\
 &                $\tilde\theta$ &  2.00 &  1.29 &  0.88 &  0.61 & 0.43 & 0.30 &  0.21 \\ \midrule
\multirow{5}{*}{PU} &                $\hat\theta$ &  0.44 &  0.46 &  0.47 &  0.48 & 0.48 & 0.49 &  0.49 \\
 &         $\theta^{(J)}$ &  0.64 &  0.60 &  0.56 &  0.54 & 0.53 & 0.52 &  0.51 \\
 & $\theta^{(B)}$ &  0.62 &  0.58 &  0.56 &  0.54 & 0.53 & 0.52 &  0.51 \\
 &                $\theta^\dagger$ &  0.53 &  0.54 &  0.54 &  0.54 & 0.53 & 0.52 &  0.51 \\
 &                $\tilde\theta$ &  0.56 &  0.56 &  0.55 &  0.54 & 0.53 & 0.52 &  0.51 \\
        \bottomrule
      \end{tabular}
    \end{center}
    \label{ratio_simulation}
  \end{table}

  Implicit RB$M$-estimation has the side-effect of
  producing an estimator that is more robust to small values of $s_X$
  than the standard $M$-estimator and the explicit RB$M$-estimator
  are. In particular, as $s_X$ becomes smaller in absolute value,
  $\hat\theta$ and $\theta^\dagger$ diverge, while $\tilde{\theta}$
  converges to $s_{XY}/s_{XX}$, which is the slope of the regression
  line through the origin of $y$ on $x$. As a result, when $\mu_X$ is
  small in absolute value, $\tilde\theta$ has not only smaller bias,
  as granted by the developments in the current paper, but also
  smaller variance than $\hat\theta$, and, hence, smaller mean squared
  error. 
  
  To illustrate the performance of the implicit and explicit
  RB$M$-estimators of the ratio $\theta$ we assume that $(X_1, Y_1)$,
  $\ldots$, $(X_n, Y_n)$ are independent random vectors from a
  bivariate distribution constructed through a Gaussian copula with
  correlation $0.5$, to have an exponential marginal with rate $1/2$
  for $X_i$ and a normal marginal for $Y_i$ with mean $10$ and
  variance $1$. Hence, $\bar{\theta} = 5$. For each
  $n \in \{10, 20, 40, 80, 160, 320\}$, we simulate
  $N_n = 250 \times 2^{16} / n$ samples. Calibrating the simulation
  size in this way guarantees a fixed simulation error for the
  simulation-based estimate of the bias of the $M$-estimator. For each
  sample, we estimate $\theta$ using the $M$-estimator $\hat\theta$,
  the implicit and explicit RB$M$-estimators $\tilde\theta$ and
  $\theta^\dagger$, respectively, the jackknife estimator
  $\theta^{(J)} = n \hat\theta - (n - 1) \sum_{i = 1}^n
  \hat\theta_{(-i)} / n$, where
  $\hat\theta_{(-i)} = \sum_{j \ne i} Y_i/\sum_{j \ne i} X_i$, and the
  ``better'' nonparametric bootstrap ratio estimator $\theta^{(B)}$ in
  \citet[Section~10.4]{efron+tibshirani:1993} using $500$
  resamples. Note here that both vanilla asymptotic bias correction
  \citep{efron:1975} and the bias-reducing adjusted score function in
  \citet{firth:1993} require expectations of products of
  log-likelihood derivatives, which require specifying the bivariate
  distribution for $(X_i, Y_i)$. Despite the simplicity of the current
  setting, even if one could confidently specify that distribution,
  the required expectations involve non-trivial analytic calculations
  and may not even be available in closed-form.

  Table~\ref{ratio_simulation} shows the simulation-based estimates of
  the bias $\expect_G\left(\theta_n - \bar\theta\right)$, mean squared
  error $\expect_G\left\{(\theta_n - \bar\theta)^2\right\}$, mean
  absolute deviation
  $\expect_G\left(\left|\theta_n - \bar\theta\right|\right)$, and
  probability of underestimation $P_G(\theta_n < \bar\theta)$ for
  $\theta_n$ being $\hat\theta$, $\theta^{(J)}$, $\theta^{(B)}$,
  $\theta^\dagger$ and $\tilde\theta$. As with the jackknife and
  bootstrap, explicit and implicit RB$M$-estimation result in a marked
  reduction of the bias.  The estimated slopes of the regression lines
  of the logarithm of the absolute value of the estimated biases for
  $\hat\theta$, $\theta^{(J)}$, $\theta^{(B)}$, $\theta^\dagger$, and
  $\tilde\theta$ on the values of $\log n$ are $-1.039$, $-1.696$,
  $-1.413$, $-1.856$, and $-2.118$, respectively, which are in
  agreement to the theoretical slopes of $-1$, $-3/2$, $-3/2$, $-3/2$,
  and $-3/2$.
  
  Reduction of the bias in this setting leads also in marked reduction
  in mean squared error and mean absolute deviation, with
  $\tilde\theta$, $\theta^\dagger$, $\theta^{(J)}$ and $\theta^{(B)}$
  performing similarly for $n \ge 10$, and markedly better than
  $\hat\theta$. The large mean squared errors of $\theta^{(J)}$ and
  $\theta^{(B)}$ for $n = 5$ are due to extreme ratio estimates
  appearing with positive probability under the resampling
  distribution. The RB$M$-estimators also have probability of
  underestimation closer to $0.5$, hence they are closer to being
  median unbiased than $\hat\theta$, $\theta^{(J)}$ and $\theta^{(B)}$
  are.

\end{example}

\section{Asymptotic distribution of reduced-bias $M$-estimators}
\label{sec:inference}

The stochastic Taylor expansion~\eqref{taylor} in the Supplementary
Material document implies that
$\hat{{\bb \theta}} - \bar{{\bb \theta}}$ and the RB$M$-estimation
counterparts $\tilde{{\bb \theta}} - \bar{{\bb \theta}}$,
${\bb \theta}^\dagger - \bar{{\bb \theta}}$ have exactly the same
$O_p(n^{-1/2})$ term in their expansions because
${\bb A}({\bb \theta}) = O_p(1)$. Hence, the argument in
\citet[Section~2]{stefanski+boos:2002} applied to that first term
gives that the implicit RB$M$-estimator $\tilde{\bb \theta}$ is such that
\begin{equation}
  \label{asymptotic_normality}
  {\bb Q}(\bar{\bb \theta})^{1/2} (\tilde{\bb \theta} - \bar{\bb \theta}) \stackrel{d}{\longrightarrow} N_p({\bb 0}_p, {\bb I}_p) \, ,
\end{equation}
and the same holds for the $M$-estimator and the explicit
RB$M$-estimator $\bb \theta^\dagger$. In~(\ref{asymptotic_normality}),
${\bb I}_p$ is the $p \times p$ identity matrix,
${\bb Q}({\bb \theta}) = {{\bb V}({\bb \theta})}^{-1}$, where
${\bb V}({\bb \theta}) = {\bb B}({\bb \theta})^{-1} {\bb M}({\bb
  \theta}) \{{\bb B}({\bb \theta})^{-1}\}^\top$, with
${\bb M}({\bb \theta}) = \expect_G[ \sum_{i=1}^k \sum_{j=1}^k {\bb
  \psi^i}({\bb \theta})\{ {\bb \psi}^j({\bb \theta})\}^\top ]$, and
${\bb B}({\bb \theta})$ is a $p \times p$ matrix with $r$th row
$-\sum_{i=1}^k \expect_G\{\nabla \psi_r^i({\bb \theta}) \}$.

An implication of~\eqref{asymptotic_normality} is that
$\hat{{\bb V}}({\bb \theta}) = {\bb j}({\bb \theta})^{-1} {\bb e}({\bb
  \theta}) \left\{{\bb j}({\bb \theta})^{-1}\right\}^\top$ evaluated
at ${\bb \theta}=\tilde{\bb \theta}$ (or
${\bb \theta}=\bb \theta^\dagger$) is a consistent estimator of the
variance-covariance matrix of $\tilde{\bb \theta}$ (or
$\bb \theta^\dagger$). This is exactly as in the case where
$\hat{{\bb V}}(\hat{\bb \theta})$ is used as an estimator of the
variance-covariance matrix of $\hat{\bb \theta}$ in the framework of
$M$-estimation \citep[see, for
example,][Section~2]{stefanski+boos:2002}. The expression for
$\hat{{\bb V}}({\bb \theta})$ appears unaltered in the second term of
the right-hand-side of
expression~\eqref{general_empirical_adjustment_matrix} for the
empirical bias-reducing adjustment. As a result, the value of
$\hat{\bb V}(\tilde{\bb \theta})$ or $\hat{\bb V}(\bb \theta^\dagger)$
and that of estimated standard errors for the estimators are available
at the final quasi Newton-Raphson update for computing the
RB$M$-estimates, as detailed in Section~\ref{sec:implementation} of
the Supplementary Material document.

In addition, if the model is correctly specified, and %
$l({\bb \theta})$ in~\eqref{objective_function} is the log-likelihood,
then the second Bartlett identity gives
${\bb M}({\bb \theta}) = {\bb B}({\bb \theta})$, which implies that
${\bb Q}({\bb \theta})$ is the expected information matrix. As a
result, the RB$M$-estimator is asymptotically efficient, exactly as
the ML estimator and the reduced-bias estimator in
\citet{firth:1993} are.
 
\begin{example}
  {\bf Ratio of two means (continued)}
  \label{ratio_estimation2}
  Table~\ref{ratio_inference} shows the estimates of the actual
  variances of $\hat\theta$ and $\tilde\theta$ for
  $n \in \{10, 20, 40, 80, 160, 320\}$ and the estimate of the mean of
  $\hat{V}(\hat\theta)$ and $\hat{V}(\tilde\theta)$, respectively. As
  expected by the arguments above, the large sample approximations to
  the variance of the estimator converge to the actual variances as
  the sample size increases.
  
  \begin{table}[t]
    \caption{Simulation-based estimates of the variances of the
      $M$-estimator $\hat\theta$ and the RB$M$-estimator
      $\tilde\theta$ ($\var_G(\hat\theta)$ and $\var_G(\tilde\theta)$,
      respectively) and of the mean of $\hat{V}(\hat\theta)$ and
      $\hat{V}(\tilde\theta)$ ($\expect_G\{\hat{V}(\hat\theta)\}$ and
      $\expect_G\{\hat{V}(\tilde\theta)\}$, respectively), from the
      simulation study of Example~\ref{ratio_estimation}. See, also
      Table~\ref{ratio_simulation}.}
    \begin{center}
      \begin{tabular}{lrrrrrr}
        \toprule
        & \multicolumn{6}{c}{$n$} \\
        \cmidrule{2-7}
        \multicolumn{1}{c}{} & \multicolumn{1}{c}{10} & \multicolumn{1}{c}{20} & \multicolumn{1}{c}{40} & \multicolumn{1}{c}{80} & \multicolumn{1}{c}{160} & \multicolumn{1}{c}{320} \\
\midrule
    $\var_G(\hat\theta)$ & 3.49 & 1.41 & 0.64 & 0.30 & 0.15 & 0.07 \\
    $\expect_G\{\hat{V}(\hat\theta)\}$ & 2.55 & 1.20 & 0.59 & 0.29 & 0.14 & 0.07 \\ \midrule
    $\var_G(\tilde\theta)$ & 3.09 & 1.31 & 0.61 & 0.30 & 0.15 & 0.07 \\
 $\expect_G\{\hat{V}(\tilde\theta)\}$  & 2.16 & 1.10 & 0.56 & 0.28 & 0.14 & 0.07 \\
\bottomrule
      \end{tabular}
    \end{center}
    \label{ratio_inference}
  \end{table}
\end{example}

Another implication of~\eqref{asymptotic_normality} is that
asymptotically valid inferential procedures, like hypothesis tests and
confidence regions for the model parameters, can be constructed based
on Wald-type and generalized score approximate pivots of the form
\begin{align}
  \label{pivots}
  W_{(e)}({\bb \theta}) & = (\tilde{\bb \theta} - {\bb \theta})^\top \left\{\hat{{\bb V}}(\tilde{\bb \theta}) \right\}^{-1}(\tilde{\bb \theta} - {\bb \theta}) \, , \\ \notag
  W_{(s)}({\bb \theta}) & = \left\{\sum_{i = 1}^k{\bb \psi}^i({\bb \theta}) + {\bb A}({\bb \theta})\right\}^\top  \left\{{\bb e}(\tilde{\bb \theta})\right\}^{-1}  \left\{\sum_{i = 1}^k{\bb \psi}^i({\bb \theta}) + {\bb A}({\bb \theta})\right\}\,,
\end{align}
respectively, which, asymptotically, have a $\chi^2_p$
distribution. The same holds when ${\bb \theta}^\dagger$ is used in
place of $\tilde{\bb \theta}$. These pivots are direct extensions of
the Wald-type and generalized score pivots, respectively, that are
typically used in $M$-estimation; see \citet{boos:1992} for discussion
about generalized score tests in $M$-estimation.

\section{Bias-reducing penalties to objective functions}
\label{sec:penalties}
\subsection{Penalized objectives}
\label{sec:penalty}

When $M$-estimation is through the maximization of an objective
function of the form~\eqref{objective_function}, implicit
RB$M$-estimation is
equivalent to the maximization of the penalized objective function
\begin{equation}
  \label{pen_loglik}
  l({\bb \theta}) - \frac{1}{2}\trace \left\{{\bb j}({\bb \theta})^{-1} {\bb e}({\bb \theta}) \right\} \, ,
\end{equation} 
assuming that the maximum exists. This equivalence can be proved by
noting that ${\bb j}({\bb \theta})$ is a symmetric matrix, and
differentiating the penalty in~\eqref{pen_loglik} to get
adjustment~\eqref{general_empirical_adjustment_matrix}.

\begin{example} \label{glm_example} {\bf Generalized linear models}
  Suppose that the distribution of $Y_i \mid {\bb x}_i$ is
  fully-specified to be exponential family with mean
  $\mu_i = h(\eta_i)$ for a known inverse link function $h(\cdot)$, with
  $\eta_i = {\bb x}_i^\top {\bb \beta}$, and variance
  $\phi v(\mu_i) /m_i$. Then, the $i$th log-likelihood contribution is
\begin{equation}
  \label{loglikelihood_glm}
  \log f_i(y_i \mid {\bb x}_i, {\bb \beta}, \phi) = \frac{m_i}{\phi} \left\{ y_i \theta_i - \kappa(\theta_i) - c_1(y_i) \right\} - \frac{1}{2} a\left(-\frac{m_i}{\phi}\right) \, ,
\end{equation}
for sufficiently smooth functions $\kappa(\cdot)$, $c_1(\cdot)$, and
$a(\cdot)$, and known prior weights $m_1, \ldots, m_n$ with $m_i > 0$,
where $\mu_i = d \kappa(\theta_i) / d\theta_i$ and
$v(\mu_i) = d^2 \kappa(\theta_i) / d
\theta_i^2$. Section~\ref{glm_appendix} in the Supplementary Materials
document gives the form of ${\bb j}({\bb \beta}, \phi)$ and
${\bb e}({\bb \beta}, \phi)$ for GLMs with
unknown dispersion parameter $\phi$.
If the dispersion parameter $\phi$ is known, as is, for example, for
the binomial and Poisson distributions, then the bias-reducing penalty
in~\eqref{pen_loglik} involves only the $({\bb \beta}, {\bb \beta})$
blocks ${\bb j}_{{\bb\beta}{\bb\beta}}({\bb \beta})$ and
${\bb e}_{{\bb\beta}{\bb\beta}}({\bb \beta})$ of
${\bb j}({\bb \beta}, \phi)$ and ${\bb e}({\bb \beta}, \phi)$,
respectively.  Simple algebra shows that the RB$M$-estimator of
${\bb \beta}$ results as the maximizer of the bias-reducing penalized
log-likelihood
\begin{equation}
  \label{pen_loglik_glm}
  \sum_{i = 1}^n m_i \left\{ y_i \theta_i - \kappa(\theta_i) - \frac{1}{2} s_i \frac{d_i}{v_i}(y_i - \mu_i) \right\}  \, ,
\end{equation}
where $d_i = d h(\eta_i) / d\eta_i$, and $v_i = v(\mu_i)$. The
quantity $s_i$ is the $i$th diagonal element of
${\bb X}({\bb X}^\top {\bb Q} {\bb X})^{-1} {\bb X}^\top \tilde{{\bb
    W}}$, where ${\bb X}$ is the $n \times p$ model matrix with rows
${\bb x}_1, \ldots, {\bb x}_n$, and the diagonal matrices
$\tilde{{\bb W}}$ and ${\bb Q}$ are as defined in
Section~\ref{glm_appendix} of the Supplementary Material.

Expression~(\ref{pen_loglik_glm}) is in contrast to the method in
\citet{firth:1993}, for which a penalized log-likelihood does not
always exist; see, \citet[Theorem 1]{kosmidis+firth:2009} for the
necessary and sufficient condition for the existence of a
bias-reducing penalized likelihood for GLMs.

\end{example}

\subsection{Estimation}
\label{sec:penalties+estimation}
The implicit RB$M$-estimates can be computed using general numerical
optimization procedures for the maximization of the penalized
objective function~\eqref{pen_loglik} that operate by numerically
approximating gradients,
like those provided by the \texttt{optim} function in R or the
\texttt{Optim} Julia package \citep{optim}. In such cases, RB$M$-estimation requires
only routines for matrix multiplication and inversion, and the
contributions to the objective function and their first two
derivatives for the implementation of ${\bb j}({\bb \theta})$ and
${\bb e}({\bb \theta})$, which can be obtained using automatic
differentiation. The \texttt{MEstimation} Julia package provides a
proof-of-concept of an implementation relying on automatic differentiation.

The explicit RB$M$-estimator
${\bb \theta}^\dagger = \hat{\bb \theta} + {\bb j}(\hat{\bb
  \theta})^{-1} {\bb A}(\hat{\bb \theta})$ in~\eqref{bias_correction}
can be obtained by computing ${\bb j}(\hat{\bb \theta})$, and
numerically differentiating the penalty term
$- \trace \left\{{\bb j}({\bb \theta})^{-1} {\bb e}({\bb \theta})
\right\} / 2$ in~\eqref{pen_loglik} at the $M$-estimate
$\hat{\bb\theta}$ to get the value of ${\bb A}(\hat{\bb \theta})$.

\begin{example}
  \label{max_stable_ex}
  {\bf Gaussian max-stable processes} Vanilla likelihood and Bayesian
  approaches for spatial extreme processes face challenges, because
  the direct generalization of the classical multivariate extreme
  value distributions to the spatial case is a max-stable process for
  which the evaluation of the likelihood becomes increasingly more
  intractable as the number of site locations increases
  \citep{davison+gholamrezaee:2012}. Several works have proposed the
  use of computationally appealing surrogates to the likelihood, like
  composite likelihoods, which are formed by specifying marginal or
  conditional densities for subsets of site locations
  \citep[see][among others]{padoan+ribadet+sisson:2010,
    genton+ma+sang:2011, davison+gholamrezaee:2012,
    huser+davison:2013}. Nevertheless, standard bias-reduction
  methods, like the one in \citet{firth:1993} and \citet{kuk:1995} in
  Table~\ref{br_characteristics}, are either infeasible or
  computationally expensive because the calculation of the bias
  function involves either integrals with respect to the true
  underlying joint density or requires repeated sampling and
  refitting.

  An example of a surrogate to the likelihood is the pairwise
  likelihood introduced in \citet{padoan+ribadet+sisson:2010} under a
  block maxima approach for modelling extremes. Suppose that
  $y_1(\bb s),\ldots,y_k(\bb s)$ with
  $\bb s \in \{\bb s_1,\ldots,\bb s_L\}$, $\bb s_j \in \Re^2$, are $k$
  independent observations at each of $L$ site locations. The pairwise
  log-likelihood formed from the collection of $L(L-1)/2$ distinct
  pairs of locations is
\begin{equation}
  \label{pwl_mev}
l({\bb \theta})=\sum_{i=1}^k\sum_{l>m} \log f(y_i(\bb s_l),y_i(\bb s_{m}) \mid {\bb \theta}) \,,
\end{equation}
where $f(y_i(\bb s_l),y_i(\bb s_{m}) \mid {\bb \theta})$ is the joint
density of $Y_i(\bb s_l)$ and $Y_i(\bb s_{m})$
$(l, m = 1, \ldots, L; l \ne m)$, given in
expression~(\ref{dens_gaussian_max}) of the Supplementary Material
document. The spatial dependence between $Y_i(\bb s_l)$ and
$Y_i(\bb s_{m})$ is characterized by the $2\times 2$ matrix
${\bb \Sigma}({\bb \theta})$ with diagonal elements $\sigma^2_1$ and
$\sigma^2_2$, and $\sigma_{12}^2$ in the off-diagonals. The maximizer
of \eqref{pwl_mev} with respect to
${\bb \theta}=(\sigma^2_{1}, \sigma^2_{2}, \sigma^2_{12})^\top$, is
the maximum pairwise likelihood estimator $\hat{\bb\theta}$, and the
RB$M$-estimator $\tilde{\bb\theta}$
maximizes~\eqref{pen_loglik}. Expressions for ${\bb j}({\bb \theta})$
and ${\bb e}({\bb \theta})$ are given in
Section~\ref{sec:gaussian_max} of the Supplementary Material document.

Simulations are run by generating independent observations
$y_1(\bb s),\ldots,y_k(\bb s)$ from a Gaussian max-stable process
observed at $L=50$ site locations. The locations are generated
uniformly on a $[0, 40] \times [0, 40]$ region.  We consider sample
sizes $k \in \{10, 20, 40, 80, 160\}$ with corresponding number of
simulations equal to $4000 k$, and true parameter settings
$\bar{\bb \theta} = (2000, 3000, 1500)^\top$ and
$\bar{\bb \theta} = (20, 30, 15)^\top$, imposing strong and weak
spatial dependence, respectively. These parameter values correspond to
${\bb \Sigma}_4$ and ${\bb \Sigma}_5$ in Table 1 of
\citet{padoan+ribadet+sisson:2010}. In Figure~\ref{MEV_plots}, we show
the simulation-based estimates of the logarithms of the absolute
biases
as functions of $\log n$, where $n = k$. These curves have roughly
slopes $-1$, and between $-3/2$ and $-2$, respectively as expected by
the asymptotic theory in Section~\ref{sec:theory}, demonstrating the
reduction of the bias that $\tilde{\bb \theta}$ delivers. Furthermore,
$\tilde{\bb \theta}$ appears to have smaller finite-sample mean
squared error, and hence smaller variance than the maximum pairwise
likelihood estimator $\hat{\bb \theta}$. The simulation results
provide evidence for the superiority of the RB$M$-estimator.

We are not including simulation- or resampling-based methods for bias
reduction (see Table~\ref{br_characteristics}) in the simulation
experiments here because they require the repeated calculation of
maximum pairwise likelihood estimates, and hence are expensive
computationally. Furthermore, the bias reduction method in
\citet{firth:1993} does not apply here because its application
requires expectations of products of derivatives of the full
likelihood function for the Gaussian max-stable process, which poses
more severe tractability issues than ML.

\begin{figure}[t!]
{\centering
{\includegraphics[width = 0.24\textwidth]{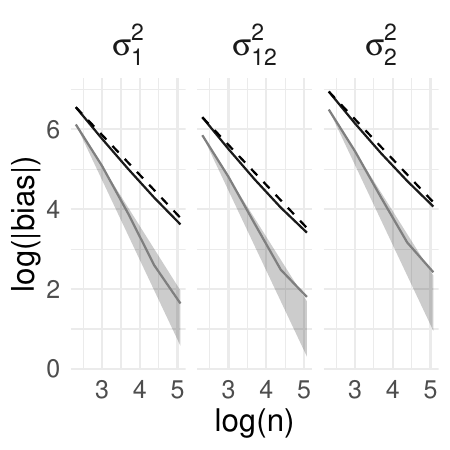}}
{\includegraphics[width = 0.24\textwidth]{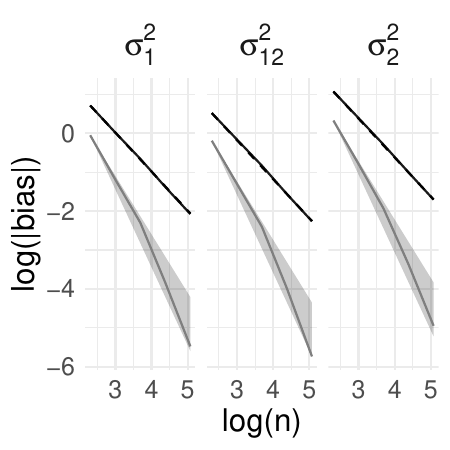}}
{\includegraphics[width = 0.24\textwidth]{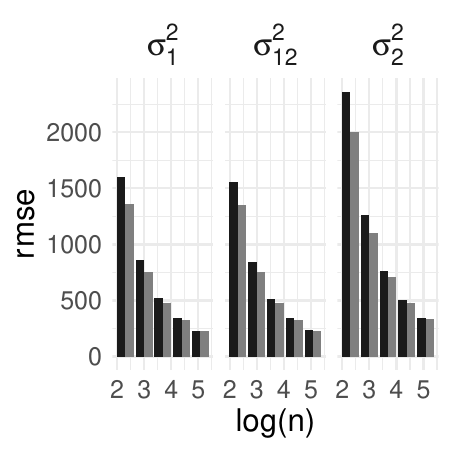}}
{\includegraphics[width = 0.24\textwidth]{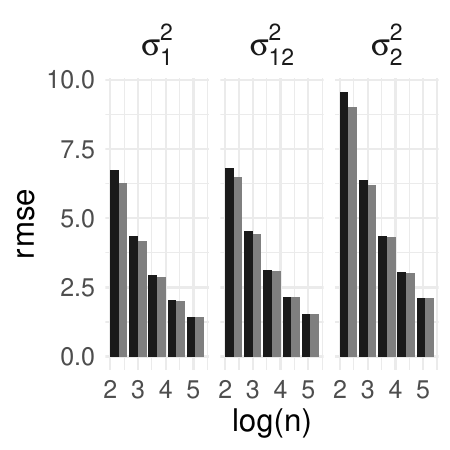}}}
\caption{The two left-most panels refer to ${\bb \Sigma}_4$ and
  ${\bb \Sigma}_5$, respectively, and show simulation-based estimates
  of the logarithm of the absolute bias of $\hat{\bb \theta}$ (black)
  and $\tilde{\bb \theta}$ (grey) from the experiments in
  Example~\ref{max_stable_ex}. The dashed line has slope $-1$
  corresponding to the theoretical rate of the bias of
  $\hat{\bb \theta}$, and the shadowed region is defined by lines with
  slopes $-3/2$ and $-2$, corresponding to the theoretical rate of the
  bias of $\tilde{\bb \theta}$. The two right-most panels show
  simulation-based estimates of the root mean squared error of
  $\hat{\bb \theta}$ (black) and $\tilde{\bb \theta}$ (grey) at
  ${\bb \Sigma}_4$ and ${\bb \Sigma}_5$.}
\label{MEV_plots}
\end{figure}

\end{example}

\subsection{Plug-in penalties and finite estimates in binomial-response GLMs}
\label{sec:plugin}

The family of bias-reducing adjustments to the estimating functions
defined by expression~\eqref{adjustment} allows for some creativity on
the construction of adjustments. For example, when $M$-estimation is
through an objective function, then maximization of~\eqref{pen_loglik}
after adding a plug-in penalty $P(\bb \theta)$ with
$\nabla P(\bb \theta) = O_p(n^{-1/2})$ still results in estimators
with bias of order $O(n^{-3/2})$. In this way, we can construct
RB$M$-estimators with extra properties (RB$M$p-estimators).

\citet{firth:1993} showed that maximizing the logistic regression
likelihood after penalizing it by the Jeffreys' invariant prior
results in reduced-bias estimates. As proved in
\citet{kosmidis+firth:2021}, such penalization has the useful
side-effect that the reduced-bias estimates are always finite, even in
cases where the ML estimates are infinite. The
finiteness property of the reduced-bias estimator is rather attractive
for applied work, because it circumvents all the numerical and
inferential issues that ML encounters with separated
datasets \citep[see,][for a recent review of the practical issues
associated with infinite estimates in logistic
regression]{mansournia+etal:2018}. In \citet{kosmidis+firth:2021}, it
is also shown that the finiteness of the maximum penalized likelihood
estimates extends more generally to penalties that are any positive
power of the Jeffreys' prior penalty, and to binomial-response
GLMs with links other than the logistic. The
resulting estimators, though, do not necessarily have better bias
properties than the ML estimator, and that bias can be
considerable for large effects; for example,
\citet[Section~8.1]{cordeiro+mccullagh:1991} show that, in logistic
regressions, the first term in the bias expansion of the maximum
likelihood estimator grows roughly linearly with the effect size.

To this end, we can appropriately select a plug-in penalty
$P(\bb \theta)$ to~\eqref{pen_loglik} and construct finite
RB$M$p-estimators for a wide range of binomial-response generalized
linear models, including logistic, probit, complementary log-log, and
cauchit regression. Consider a binomial-response GLM with $i$th likelihood contribution
$f_i(y_i \mid {\bb x}_i, {\bb \beta})$ as in~\eqref{loglikelihood_glm}
and binomial totals $m_1, \ldots, m_n$. Consider also that estimation
is by maximization of the penalized log-likelihood
  \begin{equation}
    \label{pen_loglik_plugin}
    \sum_{i = 1}^n\log f_i(y_i \mid {\bb x}_i, {\bb \beta}) - \frac{1}{2}\trace \left\{{\bb j}_{{\bb\beta}{\bb\beta}}({\bb \beta})^{-1} {\bb e}_{{\bb\beta}{\bb\beta}}({\bb \beta}) \right\} + \frac{1}{N} \log | {\bb X}^\top {\bb W}(\bb \beta) {\bb X} |  \, ,
  \end{equation}
  for some $N > 0$, where ${\bb X}$ is the $n \times p$ matrix with
  rows ${\bb x}_1, \ldots, {\bb x}_n$, assumed to have full rank, and
  ${\bb W}(\bb \beta)$ is a diagonal matrix with $i$th diagonal
  element the working weight $m_i \omega(\eta_i)$ $(i = 1, \ldots, n)$
  with $\omega(\eta) = h'(\eta)^2/[h(\eta)\{1 - h(\eta)\}]$ and
  $h'(\eta) = d h(\eta)/d\eta$.  The first two terms
  in~\eqref{pen_loglik_plugin} form the bias-reducing penalized
  log-likelihood for GLMs with known dispersion
  derived in Example~\ref{glm_example}.
  What is important to note here is that, directly by its definition,
  ${\bb e}_{{\bb\beta}{\bb\beta}}({\bb \beta})$ is symmetric and
  positive semi-definite. Furthermore, as shown in \citet{pratt:1981},
  the log-likelihood for binomial generalized-linear models is concave
  for many of the commonly used link functions, including logit,
  probit, complementary log-log, and cauchit links. Hence, for those
  link functions ${\bb j}_{{\bb\beta}{\bb\beta}}({\bb \beta})$, apart
  from being symmetric, is also positive semi-definite.  As a result
  $\trace \left\{{\bb j}_{{\bb\beta}{\bb\beta}}({\bb \beta})^{-1} {\bb
      e}_{{\bb\beta}{\bb\beta}}({\bb \beta}) \right\} \ge 0$, and the
  bias-reducing penalized log-likelihood~\eqref{pen_loglik} is bounded
  above by zero.  For the aforementioned link functions, it also holds
  that $\omega(\eta)$ converges to zero as $\eta$ diverges to either
  $-\infty$ or $+\infty$. According to \citet[Theorem~1 and
  Section~3.1]{kosmidis+firth:2021} the plug-in penalty
  $\log | {\bb X}^\top {\bb W}(\bb \beta) {\bb X} | / N$ diverges to
  $-\infty$ as any element of $\bb \beta$ diverges. Hence, the
  maximizer of~\eqref{pen_loglik_plugin} has all of its components
  finite for any $N > 0$. Choosing, for example,
  $N = \sqrt{\sum_{i = 1}^nm_i}$ or $N = \sum_{i = 1}^nm_i$ also
  guarantees that the maximizer of~\eqref{pen_loglik_plugin}, apart
  from finite components, also has bias that is free from the
  first-order term. Example~\ref{probit_bias} and
  Example~\ref{autologistic} below demonstrate the performance of implicit
  RB$M$-estimators with plug-in penalties in fully-specified probit
  regression, and partially-specified autologistic regression.
  
  \begin{example}
    \label{probit_bias} {\bf Probit regression} The performance of the
    explicit and implicit RB$M$ estimators is assessed here in a
    fully-specified, Bernoulli-response GLM with
    $\mu_i = \Phi(\beta_1 + \sum_{t = 2}^5 \beta_t x_{it})$
    $(i = 1, \ldots, n)$, where $\Phi(\cdot)$ is the cumulative
    distribution function of a standard Normal distribution. The
    covariate values $x_{i2}, x_{i3}, x_{i4}, x_{i5}$
    $(i = 1, \ldots, n)$ are generated independently and independent
    of each other from a standard normal distribution, Bernoulli
    distributions with probabilities $1/4$ and $3/4$, and an
    exponential distribution with rate $1$, respectively. Note that
    the necessary and sufficient condition of
    \citet[Theorem~1]{kosmidis+firth:2009} is not satisfied for this
    model, and hence, there is no bias-reducing penalized
    log-likelihood that corresponds to the adjusted score functions of
    \citet{firth:1993}. Instead, the bias-reducing penalized
    log-likelihood~\eqref{pen_loglik_glm} and its version with a
    plug-in penalty in~\eqref{pen_loglik_plugin} are well-defined.

    For $n \in \{100, 150, 200, 250, 300\}$, we simulate $n$ covariate
    values, as detailed in the previous paragraph, and conditional on
    those we simulate $1\, 000$ samples of $n$ response values from
    Bernoulli distributions with probabilities prescribed by a probit
    regression model with ${\bb \beta} = (-2, 2, 2, -0.5,
    -0.5)^\top$. For each sample, we estimate ${\bb \beta}$ using ML,
    explicit and implicit RB$M$-estimation, implicit RB$M$-estimation
    with plug-in penalty with $N = n$, and the adjusted score
    functions approach in \citet{firth:1993}, which relies on
    expectations of products of log-likelihood derivatives with
    respect to the correct model.

\begin{figure}[t!]
   {\centering \includegraphics[width = \textwidth]{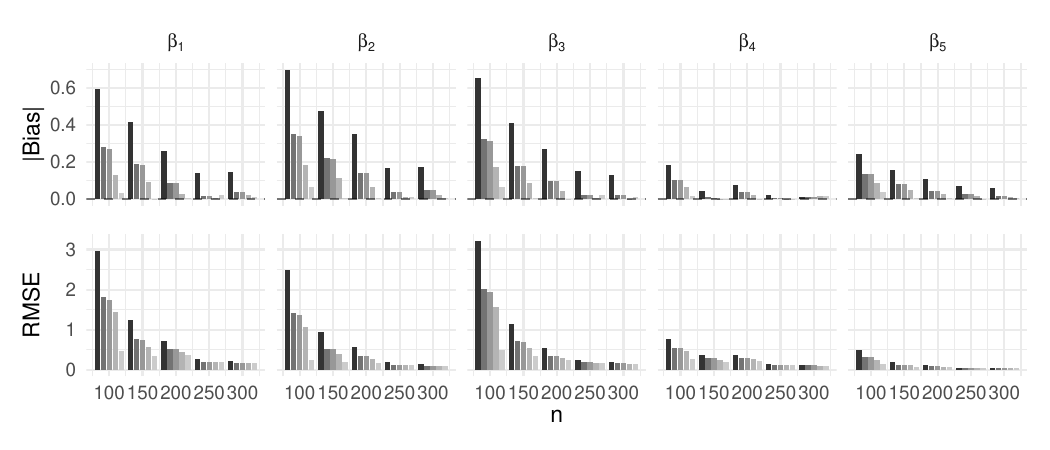}}
   \caption{Absolute bias ($|$Bias$|$) and empirical root mean squared
     error (RMSE) of various estimators of $\beta_1, \ldots, \beta_5$
     for $n \in \{100, 150, 200, 250, 300\}$ from the simulation
     setting in Example~\ref{probit_bias}. Results are shown, from
     darker to lighter grey, for the ML estimator, the
     implicit RB$M$-estimator, the implicit RB$M$-estimator with
     plug-in penalty, the explicit RB$M$-estimator, and the adjusted
     score functions estimator of \citet{firth:1993}.}
   \label{probit_bias_results}
\end{figure}
    
Maximum likelihood estimates are computed using the \texttt{glm}
function in R \citep{rproject}, and the \citet{firth:1993} adjusted
scores estimates are computed using the \texttt{brglm\_fit} method
from the \texttt{brglm2} R package \citep{brglm2}. The RB$M$-estimates
with and without plug-in penalty result from the numerical
maximization of \eqref{pen_loglik_plugin} and \eqref{pen_loglik_glm},
respectively; see the R scripts for probit regression supplied in the
Supplementary Material.
    
Infinite ML estimates were observed for $15$, and $4$ samples when
$n = 100$ and $n = 150$, respectively. The detection of infinite
estimates was done using the linear programming algorithms in
\citet{konis:2007}, as implemented in the \texttt{detectseparation} R
package \citep{detectseparation}. In those cases, maximizing the
likelihood and the bias-reducing penalized
likelihood~\eqref{pen_loglik_glm} result in estimates on the boundary
of the parameter space, and also explicit RB$M$-estimates cannot be
computed. As expected by the arguments in Section~\ref{sec:plugin},
the maximization of the bias-reducing penalized likelihood with the
plug-in penalty in~\eqref{pen_loglik_plugin} always results in finite
estimates, as did the adjusted score approach of \citet{firth:1993}.
    
Figure~\ref{probit_bias_results} shows estimates of the absolute bias
and root mean squared error of the five estimators. The summaries are
conditional on the ML estimates not being on the boundary of the
parameter space, because otherwise the bias and the root mean squared
error are not formally defined for ML, explicit RB$M$ estimation, and
implicit RB$M$-estimation with no plug-in penalty.  As is apparent,
the adjusted scores approach of \citet{firth:1993}, and explicit and
implicit RB$M$-estimation, with or without plug-in penalty, result in
estimators with substantially smaller conditional bias and mean
squared error than the ML estimator. The finite-sample bias and mean
squared error of the RBM-estimators tends to be slightly larger than
that of the adjusted scores approach of \citet{firth:1993}. The
differences diminish fast as the sample size increases.
  \end{example}
  
\begin{example}
  \label{autologistic}
  {\bf Autologistic regression models} The arguments in
  Section~\ref{sec:plugin} are used here to deliver finite
  RB$M$p-estimators for autologistic regression models, which are
  typically estimated by composite likelihoods. Autologistic
  regression \citep{besag:1972, besag:1974} is an important model for
  analysing binary responses with spatial or network correlation, used
  extensively in a range of disciplines, including ecology,
  anthropology, and computer vision; see \citet{wolters:2017} for
  references. Suppose that $y({\bb s}) \in \{-1, 1\}$ is an
  observation at location
  ${\bb s} \in \mathcal{S} = \{{\bb s}_{11}, \ldots, {\bb s}_{1c_1},
  \ldots, {\bb s}_{k1}, \ldots, {\bb s}_{kc_k}\}$, which is assumed to
  be a realization of a random variable $Y(s)$ with conditional
  probability mass function
  \begin{equation}
    \label{auto_pmf}
  f(y \mid \{y(u): u \in G({\bb s})\}, \bb{x}({\bb s}), \bb{\theta}) =
  \frac{e^{y \zeta({\bb s})}}{e^{-\zeta({\bb s})} + e^{\zeta({\bb s})}} \quad \text{with} \quad  \zeta({\bb s}) =  {\bb x}({\bb s})^\top {\bb \beta} + \lambda \sum_{u \in G({\bb s})} y(u) \,, 
\end{equation}
where ${\bb x}({\bb s})$ is a $p$-vector of covariate observed at
location ${\bb s}$,
$\bb{\theta} = (\beta_1, \ldots, \beta_p, \lambda)^\top$, and
$\lambda$ is the association parameter. The mapping $G({\bb s})$ is
assumed to be known and returns the set of all locations that are
neighbours of ${\bb s}$, excluding ${\bb s}$. We allow for
$k \in \{1, 2, \ldots\}$ disconnected clusters of observations, each
with its own neighbourhood structure, by assuming
$G({\bb s}_{ij}) \cap G({\bb s}_{i'j'}) = \emptyset$, for $i \ne i'$,
$j \in \{1, \ldots, c_i\}$ and $j' \in \{1, \ldots,
c_{i'}\}$. Negative values of $\lambda$ promote discordant responses
in the neighbourhoods, while positive values promote concordant
ones. \citet{wolters:2017} shows that the $\{-1, 1\}$ coding of the
binary responses in combination with the parameterization
in~\eqref{auto_pmf} has notable advantages over alternative proposals
for autologistic regression models (for example, the traditional
specification with $\{0, 1\}$-coding in \citealt{besag:1974}, and the,
more recent, centred autologistic model with $\{0, 1\}$-coding in
\citealt{caragea+kaiser:2009}), both in terms of estimation and
interpretation of the model parameters.  What is worth noting here is
that the regression and association effects in the corresponding
$\{0, 1\}$-coding model are $2{\bb \beta}$ and $4\lambda$,
respectively.

The joint probability mass function
$f(y({\bb s}_{11}), \ldots, y({\bb s}_{kc_k}) \mid {\bb x}({\bb
  s}_{11}), \ldots, {\bb x}({\bb s}_{kc_k}), \bb{\theta})$ with full
conditionals of the form~\eqref{auto_pmf} involves a typically
intractable normalizing constant. Nevertheless, sampling from it is
possible using Gibbs sampling or perfect sampling approaches; see,
\citet{hughes+haran+caragea:2011} and \citet{wolters:2017}, and
references therein, and the \texttt{Autologistic} Julia package
\citep{autologistic}, which implements a range of sampling
techniques. So, estimation for autologistic regression models can be
performed using Monte Carlo approximations to the likelihood
function. A computationally more attractive approach is to maximize
the pseudolikelihood
\begin{equation}
  \label{besag_pseudo_lik}
  \exp\{l({\bb \theta}) \} = \prod_{i = 1}^k \prod_{j = 1}^{c_i} f(y({\bb s}_{ij}) \mid \{y(u): u \in G({\bb s}_{ij})\}, \bb{x}({\bb s}_{ij}), \bb{\theta}) \, ,
\end{equation}
with respect to ${\bb \theta}$, which, in the terminology
of~\citet{varin+reid+firth:2011}, is a composite conditional
likelihood. \citet{besag:1975} establishes the consistency of the
$M$-estimators from the maximization of $l({\bb \theta})$. A
straightforward calculation gives that
\[
  \nabla \log f(y({\bb s}_{ij}) \mid \{y(u): u \in G({\bb s}_{ij})\}, \bb{x}({\bb s}_{ij}), \bb{\theta}) = \{z({\bb s}_{ij}) - \pi({\bb s}_{ij}) \} \tilde{\bb x}({\bb s}_{ij}) \, ,
\]
where $z({\bb s}) = \{1 + y({\bb s})\}/2$, $\pi({\bb s}) = 1 / \{1 + e^{-2 \zeta({\bb s})}\}$,
and
$\tilde{\bb x}({\bb s}) = (2 {\bb x}({\bb s})^\top, 2 \sum_{u \in G({\bb s})}
y(u))^\top$.  So, the contribution to the estimating functions from
the $i$th cluster of observations is
${\bb \psi}^{i}({\bb \theta}) = \sum_{j = 1}^{c_i} \{z({\bb s}_{ij}) -
\pi({\bb s}_{ij}) \} \tilde{\bb x}({\bb s}_{ij})$, which has expectation zero
under the joint probability mass function.
The bias-reducing penalized pseudolikelihood is
$l^{\rm (RBM)}({\bb \theta}) = l(\bb \theta) - \trace \{ {\bb j}({\bb
  \theta})^{-1} {\bb e}({\bb \theta}) \} / 2$ 
in~\eqref{pen_loglik}, with
${\bb e}({\bb \theta}) = \sum_{i = 1}^k {\bb \psi}^{i}({\bb \theta})
{\bb \psi}^{i}({\bb \theta})^\top$ and
${\bb j}({\bb \theta}) = \tilde{\bb X}^\top {\bb W}(\bb \theta)
\tilde{\bb X}$, where the matrix $\tilde{\bb X}$ has $(i, j)$th
element $\tilde{\bb x}({\bb s}_{ij})$ and ${\bb W}(\bb \theta)$ has
diagonal elements
$\pi({\bb s}_{11})\{1 - \pi({\bb s}_{11})\}, \ldots, \pi({\bb
  s}_{kc_k})\{1 - \pi({\bb s}_{kc_k})\}$.

As with logistic regression, the estimates from the maximization
of~\eqref{besag_pseudo_lik} can be infinite with positive
probability. This is particularly dangerous in the practice of these
models, because parametric bootstrap is, currently, the most
widely-used approach for uncertainty quantification and inference
about the model parameters.  There is positive probability that
optimization algorithms will fail or return a very large, in absolute
value, estimate, instead of plus or minus infinity, because their
convergence criteria have been satisfied. Then, naive handling of the
bootstrap estimates will return large bootstrap standard errors or
nonsensical bootstrap intervals, when the latter are simply formally
not well-defined. To the best of authors' knowledge, this is something
that has not been reported before in the autologistic regression
literature.  Because $l(\bb \theta)$ is bounded above by zero and
is concave \citep[see,][Theorem~5]{wolters:2017}, the same arguments
as in Section~\ref{sec:plugin} show that the RB$M$p-estimates from the
maximization of
$l^{\text{(RB}M\text{p)}}({\bb \theta}) = l^{\rm (RBM)}({\bb \theta}) + \log | \tilde{\bb
  X}^\top {\bb W}(\bb \theta) \tilde{\bb X} | / \sum_{i = 1}^kc_i$
have both improved bias properties and are always finite. Hence, they
are safe to use for bootstrap uncertainty quantification and
inferences.

\begin{table}[t]
  \caption{$M$- and RB$M$p-estimates and corresponding nominally
    $95\%$ normal and percentile bootstrap confidence intervals from
    an autologistic regression model for two subsets of the Gambia
    malaria survey data \citep{thompson+etal:1999}, as detailed in
    Example~\ref{autologistic}. The numbers in $[\cdot]$ are the
    number of simulated samples with at least one infinite component in
    the $M$-estimate.}
  \begin{center}
  \begin{tabular}{l
    r@{\extracolsep{0.05em}}r
    r@{\extracolsep{0.05em}}r@{\extracolsep{0.1em}}
    r@{\extracolsep{0.05em}}r
    r@{\extracolsep{0.05em}}r@{\extracolsep{0.1em}}
    r@{\extracolsep{0.05em}}r}
  \toprule
\multicolumn{1}{c}{} & \multicolumn{2}{c}{Estimates} & \multicolumn{4}{c}{Normal} &  \multicolumn{4}{c}{Percentile} \\ \midrule
    \multicolumn{11}{c}{Subset 1 (intercept and coefficient for netuse in $\arg\max l$ are $-\infty$ and $\infty$, respectively)} \\
  \midrule
                     & \multicolumn{1}{c}{M} & \multicolumn{1}{c}{RB$M$p} & \multicolumn{2}{c}{M [500]} & \multicolumn{2}{c}{RB$M$p [499]} & \multicolumn{2}{c}{M [500]} & \multicolumn{2}{c}{RB$M$p [499]} \\ \midrule
  intercept & $-11.43$ & $-2.60$ & $(-13.86,$ & $-9.58)$ & $(-5.39,$ & $-0.02)$ & $(-12.83,$ & $-8.78)$ & $(-4.99,$ & $0.50)$ \\
      age &   $0.16$ & $ 0.16$ & $  (0.02,$ & $ 0.31)$ & $ (0.03,$ & $ 0.30)$ & $  (0.02,$ & $ 0.31)$ & $ (0.03,$ & $0.30)$ \\
   netuse &  $12.24$ & $ 3.24$ & $  (9.81,$ & $14.09)$ & $ (2.66,$ & $ 3.77)$ & $ (10.84,$ & $14.88)$ & $ (2.83,$ & $3.71)$ \\
  treated &  $-0.41$ & $-0.39$ & $ (-1.01,$ & $ 0.45)$ & $(-0.92,$ & $ 0.27)$ & $ (-1.31,$ & $ 0.09)$ & $(-1.12$, & $0.05)$ \\
    green &  $-0.04$ & $-0.03$ & $ (-0.13,$ & $ 0.10)$ & $(-0.09,$ & $ 0.04)$ & $ (-0.19,$ & $ 0.01)$ & $(-0.12,$ & $0.00)$ \\
      phc &  $0.05$ & $ 0.04$ & $ (-0.86,$ & $ 0.76)$ & $(-0.51,$ & $ 0.52)$ & $ (-0.45,$ & $ 1.04)$ & $(-0.35,$ & $0.58)$ \\
  $\lambda$ & $0.01$ & $ 0.01$ & $ (-0.05,$ & $ 0.13)$ & $(-0.03,$ & $ 0.07)$ & $ (-0.12,$ & $ 0.02)$ & $(-0.06,$ & $0.03)$ \\
  \midrule
    \multicolumn{11}{c}{Subset 2 (no infinite components in $\arg\max l$)} \\
  \midrule
                     & \multicolumn{1}{c}{M} & \multicolumn{1}{c}{RB$M$p} & \multicolumn{2}{c}{M [1]} & \multicolumn{2}{c}{RB$M$p [0]} & \multicolumn{2}{c}{M [1]} & \multicolumn{2}{c}{RB$M$p [0]} \\ \midrule
  intercept & $-9.27$ & $-9.99$ & $(-41.20,$ & $16.98)$ & $(-27.23,$ & $11.31)$ & $(-30.22,$ & $26.54)$ & $(-34.65,$ & $3.38)$ \\
      age &  $0.16$ & $ 0.16$ & $  (0.04,$ & $ 0.27)$ & $  (0.03,$ & $ 0.30)$ & $  (0.05,$ & $ 0.28)$ & $  (0.03,$ & $0.30)$ \\
   netuse &  $0.16$ & $ 0.11$ & $ (-0.75,$ & $ 0.94)$ & $ (-0.49,$ & $ 0.72)$ & $ (-0.41,$ & $ 0.81)$ & $ (-0.54,$ & $0.63)$ \\
  treated &  $0.01$ & $-0.05$ & $ (-0.54,$ & $ 0.67)$ & $ (-0.62,$ & $ 0.53)$ & $ (-0.62,$ & $ 0.60)$ & $ (-0.62,$ & $0.59)$ \\
    green &  $0.21$ & $ 0.23$ & $ (-0.43,$ & $ 1.00)$ & $ (-0.29,$ & $ 0.65)$ & $ (-0.67,$ & $ 0.72)$ & $ (-0.10,$ & $0.83)$ \\
      phc &  $0.01$ & $ 0.14$ & $ (-1.10,$ & $ 1.50)$ & $ (-0.60,$ & $ 0.72)$ & $ (-1.64,$ & $ 0.47)$ & $ (-0.43,$ & $0.95)$ \\
   $\lambda$ &  $0.02$ & $ 0.02$ & $ (-0.02,$ & $ 0.08)$ & $  (0.01,$ & $ 0.04)$ & $ (-0.05,$ & $ 0.03)$ & $  (0.01,$ & $0.03)$ \\
  \bottomrule
  \end{tabular}
\end{center}
\label{tab:gambia} 
\end{table}

For illustration purposes, we focus on two subsets of the Gambia
malaria survey data \citep{thompson+etal:1999}, which is provided in
the \texttt{geoR} R package \citep{geoR}. The two subsets consist of
villages in the central (subset 1, with 270 children in 8 villages)
and south western (subset 2, with 321 children in 11 villages) region
of the Gambia, respectively.  The variable of interest is an indicator
denoting the presence (coded as $1$) or not (coded as $-1$) of malaria
in the blood sample taken from each child. The covariate information
includes: the age of the child in years (age), an indicator denoting
whether (coded as $1$) or not (coded as $0$) the child regularly
sleeps under a bed-net (netuse), an indicator denoting whether (coded
as $1$) or not (coded as $0$) the bed-net is treated (treated), a
satellite-derived measure of the greenness of vegetation in the
immediate vicinity of the village in arbitrary units (green), and an
indicator for the presence (coded as $1$) or absence (coded as $0$) of a
health centre in the village (phc).

Table~\ref{tab:gambia} shows the estimates from fitting an
autologistic regression model by the numerical maximization of $l({\bb \theta})$
and $l^{\text{(RB}M\text{p)}}({\bb \theta})$ on each of the subsets. For both
subsets, the first element of ${\bb x}({\bb s}_{ij})$ is set to $1$,
corresponding to an intercept parameter, and the remainder elements
are the covariate values for child $j$ in village $i$. The mapping
$G({\bb s})$ has been constructed under the assumption that any two children
in the data are neighbours only if they come from the same
village. All estimates from the maximization of $l({\bb \theta})$ for subset 2
have finite components, while, for subset 1, the estimates $-11.43$
and $12.24$ for the intercept and the parameter for netuse,
respectively, are in reality $-\infty$ and $\infty$ (detection of
infinite estimates took place using the \texttt{detectseparation} R package for
the logistic regression of $z({\bb s})$ on $\tilde{\bb x}({\bb s})$). Those
apparently finite values are artefacts due to the numerical
optimization procedure stopping prematurely by meeting the
optimizer's convergence criteria. In contrast, all RB$M$p-estimates
are finite and can be trusted at the reported
accuracy.

Table~\ref{tab:gambia} also reports $95\%$ bootstrap confidence
intervals using normal approximation and $95\%$ bootstrap percentile
intervals \citep[see][expression~(5.5) and expression~(5.18),
respectively]{davison+hinkley:1997}. The intervals have been computed
using parametric bootstrap of size $500$ at the reported estimates,
which is the current state of the art for autologistic regression,
without employing any special convention for the handling of any
infinite $M$-estimates that may arise in the bootstrap samples. The
effect of infinite estimates is dramatic in subset 1, where all
bootstrap samples at the $M$-estimates result in infinite
$M$-estimates for the intercept and the coefficient of netuse. The
impact is that the apparent evidence that bootstrap at the
$M$-estimates provides about the respective parameters being zero are
grossly exaggerated, simply reflecting the optimizer's stopping
criteria. Overall, the bootstrap intervals based on RB$M$p, apart from
being always well-defined tend to be shorter in length, and in some
cases result in different inferential conclusions than the
corresponding ones based on the $M$-estimates. For example, both
bootstrap intervals based on the RB$M$p estimates provide evidence of
positive association in subset 2, while the ones based on
$M$-estimates provide no evidence of association.

The bootstrap intervals based on the RB$M$p estimator are also found
to perform better in terms of coverage than the typically wider
intervals based on the $M$-estimator. We simulated 1000 samples at
RB$M$p estimates from subset 2 in Table~\ref{tab:gambia}, and for each
sample we computed the nominally $95\%$ bootstrap percentile intervals
based on the $M$- and the RB$M$p-estimators using a bootstrap of size
500. No infinite $M$-estimates were detected in the simulated
samples. The estimated coverage probabilities for $M$-estimation in
the order that the parameters appear in Table~\ref{tab:gambia} are
$0.84$, $0.95$, $0.92$, $0.92$, $0.82$, $0.81$, $0.82$,
respectively. The corresponding estimated coverage probabilities based
on RB$M$p-estimation are found to be markedly closer to the nominal
level with $0.90$, $0.95$, $0.94$, $0.94$, $0.89$, $0.91$, $0.90$,
respectively, with the intervals also having shorter lengths at
$75\%$, $99\%$, $72\%$, $17\%$, $75\%$, $18\%$, $77\%$, respectively,
of the length of the $M$-estimation ones. Properties of the various
variants of bootstrap intervals and their suitability for autologistic
regression is an interesting topic that is beyond the scope of the
current work.

Note here that the adjusted score equations of \citet{firth:1993} do
not apply easily because the joint probability mass function with full
conditionals of the form~(\ref{auto_pmf}) is typically
intractable. Furthermore, alternative bias reduction methods such
bootstrap, indirect inference, and the jackknife are not well-defined,
due to the positive probability of infinite maximum pseudo-likelihood
estimates. In fact, if infinite estimates go undetected, such methods
can easily return nonsensical estimates.

\end{example}

\subsection{Links to model selection using Kullback-Leibler divergence}
\label{sec:model_selection}
Suppose that $l({\bb \theta})$ is the log-likelihood function based on an
assumed parametric model $F$. \citet{takeuchi:1976} showed that
\begin{equation}
  \label{tic}
  - 2 l(\hat{\bb \theta}) + 2 \trace \left\{{\bb j}(\hat{\bb \theta})^{-1} {\bb e}(\hat{\bb \theta}) \right\}
\end{equation}
is an estimator of the expected Kullback-Leibler divergence of the
underlying process $G$ to the assumed model $F$, where
$\hat{\bb \theta}$ is the ML estimator.  Expression~\eqref{tic} is
known as the Takeuchi information criterion (TIC), and, in contrast to
the Akaike Information Criterion (AIC; \citealt{akaike:1974}), is
robust against deviations from the assumption that the model is
correct.  \citet[Section~2.5]{claeskens:2008} thoroughly discuss the
relationship between TIC and AIC.

Model selection from a set of parametric models proceeds by computing
$\hat{\bb \theta}$ for each model and selecting the model with the
smallest TIC value~\eqref{tic}, or equivalently, with the largest
\begin{equation}
  \label{tic2}
  l(\hat{\bb \theta}) - \trace \left\{{\bb j}(\hat{\bb \theta})^{-1} {\bb e}(\hat{\bb \theta}) \right\} \, .
\end{equation}
A direct comparison of expressions~\eqref{tic2} and~\eqref{pen_loglik}
reveals a previously unnoticed close connection between bias reduction
in ML estimation and model selection. Specifically,
both bias reduction and TIC model selection rely on exactly the same
penalty
$\trace \left\{{\bb j}({\bb \theta})^{-1} {\bb e}({\bb \theta})
\right\}$, but differ in the strength of penalization; bias reduction
is achieved by using half that penalty, while valid model selection
requires stronger penalization by using one times the penalty.

As discussed in Section~\ref{sec:inference}, the explicit and implicit
RB$M$-estimators ${\bb \theta}^\dagger$ and $\tilde{\bb \theta}$,
respectively, have the same asymptotic distribution as
$\hat{\bb \theta}$. Then, the derivation of TIC \citep[see, for
example,][Section~2.3]{claeskens:2008} works also with the
RB$M$-estimators in place of the ML estimator. As a result, TIC and,
under extra assumptions, AIC at the RB$M$-estimates are asymptotically
equivalent to their versions at the maximum likelihood estimates. The
same holds for reduced-bias estimators of \citet{firth:1993}. In other
words, TIC model selection can proceed by selecting the model with the
largest value of \begin{equation}
  \label{tic3}
  l(\tilde{\bb \theta}) - \trace \left\{{\bb j}(\tilde{\bb \theta})^{-1} {\bb e}(\tilde{\bb \theta}) \right\} \, ,
\end{equation}
and AIC model selection using the largest value of
$l(\tilde{\bb \theta}) - p$, and the same holds when
${\bb \theta}^\dagger$ is used in place of $\tilde{\bb \theta}$. The
quantity in~\eqref{tic3} is readily available once~\eqref{pen_loglik}
has been maximized to obtain the implicit RB$M$-estimates; the only
requirement for model selection is to adjust, from $1/2$ to $1$, the
factor of the value of
$\trace \left\{{\bb j}({\bb \theta})^{-1} {\bb e}({\bb \theta})
\right\}$ after maximization. The same holds when $\tilde{\bb \theta}$ is
obtained using plug-in penalties, as in~\eqref{pen_loglik_plugin}.

\citet{varin+vidoni:2005} developed a model selection procedure when
the objective $l({\bb \theta})$ is a composite likelihood
\citep[see,][for a review of composite likelihood
methods]{varin+reid+firth:2011}. The composite likelihood information
criterion (CLIC) derived in \cite{varin+vidoni:2005} has the same
functional form as TIC in~\eqref{tic}. So, the link between model
selection and bias reduction exists also when $l({\bb \theta})$ is the
logarithm of a composite likelihood. From the discussion in
Section~\ref{sec:penalties+estimation} it follows that both implicit
and explicit RB$M$-estimation are readily available when there is a
ready implementation of TIC for likelihood problems, and CLIC for
composite likelihood problems \citep[see, e.g.,][for estimation of
random fields based on composite likelihoods]{padoan+bevilacqua:2015}.

\begin{example}
  \label{probit_model_selection}
  {\bf Model selection in probit regression}
  The performance of model selection procedures is assessed here using
  the probit regression model in Example~\ref{probit_bias}.  There are
  $16$ possible nested probit regression models with an intercept
  $\beta_1$, depending on which of $\beta_2, \ldots, \beta_5$ are zero
  or non-zero.  For $n \in \{75, 150, 300, 600\}$, we simulate $n$
  covariate values, as detailed in Example~\ref{probit_bias}, and
  conditional on those we simulate $10\, 000$ samples of $n$ response
  values from Bernoulli distributions with probabilities prescribed by
  a probit regression model with
  ${\bb \beta} = (-0.5, 0, 0, 0.5, 0.5)^\top$. For each sample, we
  estimate all $16$ possible models using maximum likelihood, the
  adjusted score functions approach in \citet{firth:1993}, and by
  maximizing the bias-reducing penalized likelihoods with and without
  plug-in penalty, in~\eqref{pen_loglik_plugin}
  and~\eqref{pen_loglik_glm}, respectively, where we use $N = n$ for
  the plug-in penalty.

  There were $7$ separated data sets for at least one of the 16
  models, detected using the \texttt{detectseparation} R package.
  Figure~\ref{model_selection} shows the selection proportion among
  the 16 models based on AIC and TIC at the ML estimates, the implicit
  RB$M$-estimates, the implicit RB$M$-estimates with plug-in penalty
  with $N = n$, and the adjusted score functions estimates in
  \citet{firth:1993}. As expected by the discussion in
  Section~\ref{sec:model_selection}, the probability of selecting the
  model with $\beta_2 = \beta_3 = 0$ increases with the sample size
  for both information criteria and for all estimation methods. There
  are only small discrepancies on the selection proportions between
  estimation methods, which tend to disappear as the sample size
  increases. Finally, it is worth noting that AIC model selection
  tends to be more confident on what the true model is than TIC,
  illustrating less variability in the selected proportions. In the
  current study, for the larger samples sizes this results in
  selecting the correct model more often than TIC does. However, in
  smaller samples sizes, AIC selects the model with
  $\beta_2 = \beta_3 = \beta_4 = 0$ more often.

  \begin{figure}[t!]
    {\centering \includegraphics[width = \textwidth]{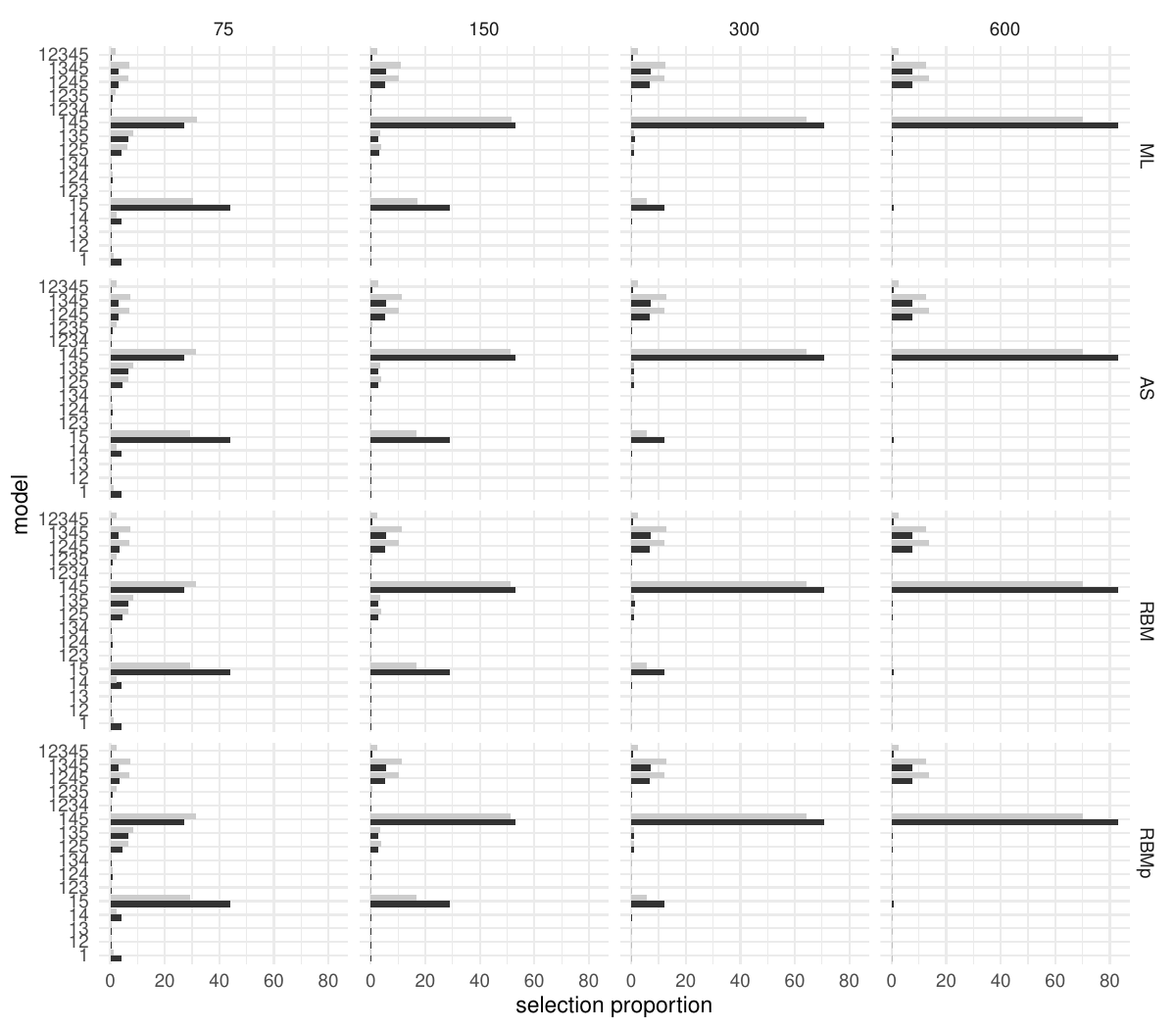}}
    \caption{Model selection proportions based on AIC (black) and TIC
      (grey) among the 16 nested models of the probit regression in
      the simulation setting of Example~\ref{probit_model_selection}
      when estimation is through ML, maximum bias-reducing penalized
      likelihood without (RB$M$) and with (RB$M$p) plug-in penalty,
      and the adjusted scores (AS) approach of \citet{firth:1993}. The
      $y$-axes give the indices of the elements of $\bb \beta$ present
      in the estimated model. }
  \label{model_selection}
\end{figure}

\end{example}

\section{High-dimensional regression settings}
\label{sec:highdim}
For the estimation of logistic regression models with
$p/n \to \kappa \in (0, 1)$, experiments reported in the supplementary
information of \citet{sur+candes:2019} and the supplementary material
of \citet{kosmidis+firth:2021} illustrate that the bias reduction
method of \citet{firth:1993} performs similarly to the methods of
\citet{sur+candes:2019}, which are based on approximate message
passing algorithms, and markedly better than ML. Reduced-bias
$M$-estimators are found here to perform well in high-dimensional,
partially-specified regression settings, where the bias-reduction
methods of \citet{firth:1993} and the methods of
\citet{sur+candes:2019} do not apply directly. We note here that the
sufficient conditions we used (see Section~\ref{sec:assumptions_sup}
of the Supplementary Material document) for developing reduced-bias
$M$-estimation do not cover for $p / n \to \kappa \in (0, 1)$.

\begin{example}
  \label{highdim_autologistic}
  {\bf High-dimensional autologistic regression} To
  illustrate, we consider the autologistic regression model
  in~(\ref{auto_pmf}), with $k = 100$ fully-connected groups, with
  $c = c_1 = \ldots = c_{100} = 10$, where each covariate vector
  $\bb{x}({\bb s}_{ij})$ has $p = 100$ entries simulated independently
  from a Normal distribution with mean $0$ and variance
  $(2kc)^{-1}$. We consider five high-dimensional
  autologistic regression models, for
  $\lambda \in \{-0.2, -0.1, 0, 0.1, 0.2\}$ and where $\bb{\beta}$ has
  the first $20$ entries having value $10$, the next $20$ entries
  having value $5$, and the remaining $60$ entries having value
  $0$. For these parameter settings, the marginal probabilities within
  each group range from spanning $(0, 1)$ ($\lambda = -0.2$, which
  promotes discordant, high-variance neighbourhoods) to being
  similarly extreme in value ($\lambda = 0.2$, which promotes
  concordant, low-variance neighbourhoods).

  \begin{table}
    \caption{Variance and the mean squared error to variance ratio (in
      parenthesis) for the $M$-estimator and the RB$M$p-estimator of
      $\bb{\beta}$ for the distinct values in the true value for
      $\bb{\beta}$ in the simulation setting of
      Example~\ref{highdim_autologistic}. $M$p is $M$-estimation
      penalized just by the plug-in penalty that ensures finiteness of
      the estimates.}
    \label{tab:auto_var}
    \begin{center}
      \begin{tabular}{D{.}{.}{2}
        D{.}{.}{2}D{.}{.}{2}D{.}{.}{2}
        D{.}{.}{2}D{.}{.}{2}D{.}{.}{2}
        D{.}{.}{2}D{.}{.}{2}D{.}{.}{2}}
        \toprule
       \multicolumn{1}{c}{$\lambda$} & \multicolumn{3}{c}{$M$-estimator} & \multicolumn{3}{c}{$M$p-estimator} & \multicolumn{3}{c}{RB$M$p-estimator} \\ \midrule
       &
      \multicolumn{1}{c}{0} & \multicolumn{1}{c}{5} & \multicolumn{1}{c}{10} &
      \multicolumn{1}{c}{0} & \multicolumn{1}{c}{5} & \multicolumn{1}{c}{10} &
      \multicolumn{1}{c}{0} & \multicolumn{1}{c}{5} & \multicolumn{1}{c}{10} \\
        \cmidrule{2-10}
-0.20 &  6.23 &  6.83 &  6.72 &  6.23 &  6.83 &  6.71 &  4.80 &  5.24 &  5.07 \\
& (1.00) & (1.15) & (1.52) & (1.00) & (1.15) & (1.51) & (1.00) & (1.01) & (1.03) \\
-0.10 &  5.80 &  6.35 &  6.24 &  5.80 &  6.35 &  6.23 &  4.51 &  4.92 &  4.75 \\
&  (1.00) & (1.14) & (1.45) & (1.00) & (1.14) & (1.45) & (1.00) & (1.01) & (1.02) \\
 0.00 &  5.37 &  5.84 &  5.78 &  5.37 &  5.84 &  5.78 &  4.36 &  4.83 &  4.87 \\
&  (1.00) & (1.12) & (1.44) & (1.00) & (1.12) & (1.44) & (1.00) & (1.01) & (1.03) \\
 0.10 &  6.21 &  6.41 &  6.31 &  6.21 &  6.40 &  6.31 &  4.79 &  4.92 &  4.80 \\
&  (1.00) & (1.15) & (1.52) & (1.00) & (1.15) & (1.52) & (1.00) & (1.01) & (1.03) \\
 0.20 & 19.94 & 21.04 & 24.85 & 19.90 & 21.00 & 24.76 & 13.27 & 13.71 & 15.61 \\
&  (1.00) & (1.21) & (1.74) & (1.00) & (1.21) & (1.74) & (1.00) & (1.04) & (1.16) \\
        \bottomrule
\end{tabular}
\end{center}
\end{table}

We simulate $100$ independent response samples at each
$\lambda \in \{-0.2, -0.1, 0, 0.1, 0.2\}$ and for each sample we
estimate ${\bb \beta}$ and $\lambda$.  Table~\ref{tab:auto_var} shows
the variance and the mean squared error to variance ratio for the
$M$-estimator and the RB$M$p-estimator of $\bb{\beta}$ for the
distinct values in the true value for $\bb{\beta}$. All estimates had
finite components. A mean squared error to variance ratio close to one
indicates that the bias has small contribution to the mean squared
error. Both $M$ and RB$M$p estimators are close to being unbiased for
the zero effects in $\bb{\beta}$. Nevertheless, the contribution of
the bias to the mean squared error of the $M$-estimator increases
rapidly as the effects sizes grow; for the entries of $\bb{\beta}$
with value $10$, the mean squared error of the $M$-estimator is as
much as $74\%$ larger than its variance. Remarkably, the
RB$M$p-estimator apart from smaller variance for all parameters has
mean squared error to variance ratios close to one. For reference,
Table~\ref{tab:auto_var} also reports figures for $M$-estimation
penalized just by the plug-in penalty. The figures are almost
identical to those for $M$-estimation indicating that the excellent
bias and variance properties the RB$M$p-estimator has in this setting
are due to bias-reducing penalty
$-\trace\{{\bb j}(\bb \theta)^{-1} {\bb e}(\bb \theta)\}/2$, with the
plug-penalty having almost no impact on the $M$-estimates.
\end{example}

\section{Correlated random components}
\label{spatiotemporal}

In Section~\ref{empirical_adjustments}, we have shown that when
${\bb y}_1, \ldots, {\bb y_k}$ are realizations of independent random
vectors ${\bb Y}_1, \ldots, {\bb Y_k}$, the implicit and explicit
RB$M$-estimation will result in estimates with superior bias
properties than the corresponding $M$-estimates. This is true
regardless of the dependence structure imposed by the model on the
components of each random vector ${\bb Y}_i$; this fact is illustrated
in Example~\ref{max_stable_ex} in the context of Gaussian max-stable
processes. However, it is often the case that $k = 1$, in the sense
that the vector of observations ${\bb y} = (y_1, \ldots, y_T)^\top$ is
assumed to be a single realization of a random vector ${\bb Y}$ with
correlated components, like when a single time series or spatial
process is observed, and the information about $\bb \theta$ is assumed
or expected to increase with $T$. In such cases, ${\bb j}(\bb \theta)$
and ${\bb u}_r(\bb \theta)$ in the empirical
adjustment~\eqref{general_empirical_adjustment_matrix} remain valid
estimators of $\mu_{rs}(\bb \theta)$ and $\mu_{rst}(\bb \theta)$
in~\eqref{adjustment}, respectively. However, ${\bb e}(\bb \theta)$
and ${\bb d}_r({\bb \theta})$ tend to be rather imprecise estimators
of $\nu_{r, s}(\bb \theta)$ and $\nu_{rs, t}(\bb \theta)$, and their
use may duly result in the original estimator $\hat{\bb \theta}$
because
${\bb e}(\hat{\bb \theta}) = {\bb d}_r(\hat{\bb \theta}) = {\bb 0}_{p
  \times p}$ when $k = 1$.

In fact, estimation of $\nu_{r, s}(\bb \theta)$ when $k = 1$ is a
topic that has attracted much research, which mainly focuses on
estimating the covariance matrix of composite likelihood estimators
\citep[see, for example][Section~5]{varin+reid+firth:2011}, or of
$M$-estimators in regression problems more generally \citep[see, for
example][]{heagerty+lumley:2000}. In such cases, weak stationarity
conditions \citep{carlstein:1986, heagerty+lumley:2000} on the model
allow the use of window sub-sampling to evaluate the estimating
functions (or derivatives of those) over multiple, overlapping or
non-overlapping subsets of the temporal or spatial domain, and
aggregate the resulting contributions. Similar procedures can be used
for the estimator of $\nu_{rs, t}(\bb \theta)$.

\begin{example}
  \label{OLS_AR}
{\bf Least squares for autoregressive processes}
Let $\{Y_1, \ldots, Y_T\}$ be an autoregressive process of order one,
in the sense that $Y_t = \theta Y_{t-1} + \epsilon_t$ with
$\theta \in (-1,1)$, where $\epsilon_1, \ldots, \epsilon_T$ are
independent and identically distributed random variables with zero
mean and finite variance. The ordinary least squares estimator of
$\theta$ is $\hat\theta = S_1(\Omega)/S_2(\Omega)$, where
$S_1(\omega) = \sum_{t \in \omega} Y_{t + 1} Y_t$,
$S_2(\omega) = \sum_{t \in \omega} Y_t^2$,
$\Omega = \{1, \ldots, T - 1\}$, and $Y_t = 0$ if
$t \notin \Omega \cup \{T\}$.  The estimator $\hat\theta$ results by
maximizing the objective
$l(\theta)=-\sum_{t \in \Omega} (Y_{t + 1}-\theta Y_t)^2$ or,
equivalently, by finding the root of the estimating function
$\psi(\theta) = 2S_1(\Omega) - 2\theta S_2(\Omega)$.

The implicit and explicit RB$M$-estimators can be computed using the
bias-reducing penalized objective in~\eqref{pen_loglik}. For this
model, $j(\theta) = 2S_2(\Omega)$, and in place of $e(\theta)$ we use
$e^{(ws)}(\theta)=\kappa_T\sum_{r=1}^R\{2S_1(\omega_r)-2\theta
S_2(\omega_r)\}^2$ where $\kappa_T > 0$ is a suitable constant. The
quantity $e^{(ws)}(\theta)$ is a window sub-sampling estimator of
$\expect_G[\{2S_1(\Omega) - 2\theta S_2(\Omega)\}^2]$
\citep{carlstein:1986}, where $\omega_1, \ldots, \omega_R$ is a
partitioning of $\Omega$, with $\omega_r$ having indices of
observations that are consecutive in time. Then, the implicit and
explicit RB$M$-estimators are
\begin{align*}
  \tilde\theta = \frac{S_{12}(\Omega) + \kappa_T S_{12}(\omega_1, \ldots, \omega_R)}{
S_{22}(\Omega) + \kappa_T S_{22}(\omega_1, \ldots, \omega_R)} \quad \text{and} \quad
  \theta^\dagger = \hat\theta\left\{1+\kappa_T\frac{S_{22}(\omega_1, \ldots, \omega_R)}{S_{22}(\Omega)}\right\} - \kappa_T\frac{S_{12}(\omega_1, \ldots, \omega_R)}{S_{22}(\Omega)}\,,
\end{align*}
respectively, where
$S_{lm}(\omega_1, \ldots, \omega_R) =
\sum_{r=1}^RS_l(\omega_r)S_m(\omega_r)$.  Note that the reduced-bias
estimator $\theta^{*}$ of \citet{firth:1993} can only be obtained
under specific assumptions about the distribution of
$\epsilon_1, \ldots, \epsilon_T$. If $\epsilon_t \sim N(0,1)$, then
the root of the adjusted score function
$2S_1(\Omega) - 2\theta S_2(\Omega) - 4(T - 1)/\{T(1 - \theta^2)\}$
has second-order bias. The latter expression is a polynomial of order
$3$ and, hence, it can have up to $3$ distinct roots, with at least
one of them being real. In contrast, $\tilde{\theta}$ and
$\theta^\dagger$ always have a single value.

\begin{table}[t]
  \caption{Simulation-based estimates of the bias ($\times 100$) of
    the various estimators for the parameter of the AR(1) process in
    Example~\ref{OLS_AR}. Figures are reported in 2 decimal places,
    and the figures $-0.00$ are for estimated biases in the interval
    $(-0.0025, -0.0025)$. The simulation error for the estimates of
    the bias is between $2.76\times 10^{-4}$ and $1.37\times
    10^{-3}$. The last two columns are the estimated (Est) and
    expected (Exp) slope of the regression line
    $\log|\text{bias}| \sim \log T$.}
  \begin{center}
    \begin{tabular}{lccD{.}{.}{2}D{.}{.}{2}D{.}{.}{2}D{.}{.}{2}D{.}{.}{2}D{.}{.}{2}r}
      \toprule
      & \multirow{2}{*}{$\alpha$} & \multirow{2}{*}{Errors} &
                                                              \multicolumn{5}{c}{$T$} & \multicolumn{2}{c}{Slope} \\ \cmidrule{4-10}
      & & & \multicolumn{1}{c}{50} & \multicolumn{1}{c}{100} & \multicolumn{1}{c}{200} & \multicolumn{1}{c}{400} & \multicolumn{1}{c}{800} &  \multicolumn{1}{c}{Est} & \multicolumn{1}{c}{Exp}\\ \midrule
      $\hat{\theta}$ & & & -1.95 & -0.98 & -0.50 & -0.23 & -0.13 & -0.99 & $-1$ \\  \cmidrule{1-10}
      \multirow{2}{*}{$\tilde{\theta}$} & $1/3$ & & -1.41 & -0.56 & -0.23 & -0.08 & -0.04 & -1.32 & $-3/2$ \\ 
      & $1/2$ & & -1.20 & -0.44 & -0.16 & -0.04 & -0.02 & -1.49 & $-3/2$ \\ \cmidrule{1-10}
      \multirow{2}{*}{$\theta^{\dagger}$} & $1/3$ & & -1.36 & -0.54 & -0.22 & -0.08 & -0.04 & -1.32 & $-3/2$ \\           & $1/2$ & & -1.07 & -0.38 & -0.14 & -0.03 & -0.02 & -1.53 & $-3/2$  \\ \cmidrule{1-10}
      $\theta^{(J)}$ & & & -0.22 & -0.04 & -0.02 & 0.02 & -0.00 & -1.468 & $<-1$ \\ \cmidrule{1-10}
      $\theta^{(M)}$ & & & -2.47 & -1.33 & -0.71 & -0.34 & -0.18 &  -0.945 & $<-1$ \\  \cmidrule{1-10}
      \multirow{2}{*}{$\theta^{(S)}$} & $1/3$ & & 10.44 & 9.28 & 8.08 & 7.08 & 5.53 & -0.22 & $<-1$  \\ 
      & $\scriptstyle\log(T/2)/\log(T)$ & & 1.33 & 0.79 & 0.43 & 0.24 & 0.11 & -0.89 & $<-1$ \\ \cmidrule{1-10}
      &  & Normal & 0.30 & 0.07 & 0.01 & 0.02 & -0.00 & -1.60 & $-2$ \\ 
      $\theta^{*}$ & & Student-t & 2.33 & 0.88 & 0.38 & 0.19 & 0.08 & -1.18 & \\
      & & Laplace & -12.55 & -8.49 & -5.68 & -3.98 & -2.75 & -0.55 & \\ 
      \bottomrule
    \end{tabular}
  \end{center}
  \label{tab:OLS}
\end{table}

We conduct a simulation study to assess the finite-sample bias
properties of $\hat\theta$, $\tilde\theta$, $\theta^\dagger$, and
$\theta^{*}$. We simulate $N_T$ independent time series of lengths
$T \in \{50,100,200,400,800\}$, with varying strength of dependence
$\theta \in \{0.2, 0.5, 0.9\}$. The simulation size is set to
$N_T = 250\times2^{16}/T$ to control the simulation error, similarly
to Example~\ref{ratio_estimation}. The errors $\epsilon_t$
$(t = 1, \ldots, T)$ are simulated so that the joint distribution of
the random vector $(Y_1, \ldots, Y_T)^\top$ is multivariate normal,
multivariate Student-t with $5$ degrees of freedom, and multivariate
asymmetric Laplace, all with mean ${\bb 0}_T$ and covariance matrix
with $(t, k)$ entry $\text{Cov}(Y_t,Y_k)=\theta^{|t-k|}$, which is the
exponential decaying function of an autoregressive process of order
one. We compute $\theta^*$ for all three error structures with the
bias-reducing adjustment computed at the model with normal errors to
illustrate the impact of using the wrong bias adjustment, and we only
keep the root of the polynomial that is returned by starting at
$\hat\theta$. In order to investigate the impact of the subsampling
scheme on $\tilde\theta$ and $\theta^\dagger$, we consider
$m_T = T^\alpha$ with the $\alpha \in \{1/3, 1/2\}$. The choice
$\alpha = 1/3$ is the one that minimizes the mean squared error of
$e^{(ws)}(\theta)$ \citep{carlstein:1986} for AR(1) models. In the
comparison, we also include the bootstrap bias-corrected estimators
$\theta^{(M)}$ using residual resampling, and $\theta^{(S)}$ using the
stationary bootstrap \citep[see][Chapter 8]{davison+hinkley:1997},
computed from 500 bootstrap samples, and as implemented in the
\texttt{tsboot} function in the \texttt{boot} R package \citep{boot}. The block
scheme for the stationary bootstrap is chosen to have average length
$T^{\alpha}$, with $\alpha \in \{1/3, \log(T/2)/\log(T)\}$. Average
block length proportional to $T^{1/3}$ is optimal under the general
guidelines in \citet{kunsch:1989} and \citet{politis+romano:1994} for the
standard error of $\hat\theta$. The other choice for $\alpha$ sets the
average block length equal to $T/2$. We also include the half sample
jackknife estimator $\theta^{(J)}$ \citep{quenouille:1949}.

Table~\ref{tab:OLS} gives the simulation-based estimates of the bias
of estimators at $\theta = 0.5$. The results for
$\theta \in \{0.2, 0.9\}$ are in the Supplementary Material. The
estimated biases $\hat\theta$, $\tilde\theta$, $\theta^\dagger$,
$\theta^{(J)}$, $\theta^{(M)}$, and $\theta^{(S)}$ in
Table~\ref{tab:OLS} are identical for all three error distributions we
consider. The reason is that by mere inspection of their expressions,
the RB$M$-estimates, like the ordinary least squares estimates and, in
turn, bootstrap- and jackknife-based versions, are invariant to
scaling the time series by a scalar random variable $C$ that is
independent to the distribution of
$(\epsilon_1, \ldots, \epsilon_T)$, and both the multivariate
Student-t and asymmetric Laplace distributions are infinite mixtures
of the multivariate Normal distribution. The estimator $\theta^*$ does
not have such invariance.

Table~\ref{tab:OLS} also reports the estimated and theoretical slopes
from the regression of the logarithm of the absolute bias to $\log
T$. We observe a close agreement between estimated and expected slopes
for $\hat\theta$, $\tilde\theta$, $\theta^\dagger$, and $\theta^*$
under Normal errors. We also note that RB$M$-estimators deliver the
bias reduction expected by the theory in Section~\ref{sec:theory} for
both window sizes considered. As the error distribution departs from
Normal, the use of an incorrect adjustment impacts $\theta^*$ both in
terms of finite sample bias and the asymptotic rates of the bias,
which drop dramatically in absolute value from what is expected under
Normal errors. The jackknife estimator $\theta^{(J)}$ is as effective
as $\tilde\theta$ and $\theta^{\dagger}$ in terms of reducing bias. On
the other hand, use of $\theta^{(M)}$ appears to have not much effect
in reducing the bias of $\hat\theta$ with an estimated slope close to
$-1$, and $\theta^{(S)}$ delivers both larger finite sample bias than
$\hat\theta$ and worse asymptotic rates for the bias.

When the ad-hoc average block length $T/2$ is used, the performance of
$\theta^{(S)}$ in terms of bias improves. Hence, $\theta^{(S)}$
appears to require a different tuning of the average block length than
$T^{1/3}$, which is optimal for standard error estimation.

\end{example}

\section{Discussion and further work}
\label{sec:discussion}

We have developed a novel and general framework for the reduction of
the asymptotic bias of general $M$-estimators from asymptotically
unbiased estimating functions. The framework relies on a novel
approximation of the bias function that requires only the
contributions to the estimating functions and the first two
derivatives of those, which we used to derive an explicit and
implicit method for reduced-bias $M$-estimation of general
applicability. Explicit RB$M$-estimation proceeds by subtracting the
approximation of the bias function at the $M$-estimates from the
$M$-estimates. Implicit RB$M$-estimates result as roots of additively
adjusted estimating functions, with the empirical bias-reducing
adjustment~\eqref{general_empirical_adjustment_matrix}. Both explicit
and implicit RB$M$-estimates can be computed using the quasi
Newton-Raphson iteration in Section~\ref{sec:implementation} of the
Supplementary Material document. The RB$M$-estimators have the same
asymptotic distribution, and, hence, they are asymptotically as
efficient as the initial $M$-estimators. As detailed in
Section~\ref{sec:inference}, uncertainty quantification can be carried
out using the empirical estimate $\hat{{\bb V}}({\bb \theta})$ of the
variance-covariance matrix of that asymptotic distribution. The
expression for $\hat{{\bb V}}({\bb \theta})$ is part of
expression~\eqref{general_empirical_adjustment_matrix} for the
empirical bias-reducing adjustment, and, hence, is readily available
at the last iteration of the quasi Newton-Raphson iterative procedure.
Inferences can be constructed using the Wald-type and generalized
score pivots in expression~\eqref{pivots}.

If $M$-estimation is by the maximization of an objective function,
then implicit RB$M$-estimation can always be achieved by the
maximization of the bias-reducing penalized
objective~\eqref{pen_loglik}, which closely relates to model selection
procedures using the Kullback-Leibler divergence; the functions of the
parameters and the data that are used for bias reduction and model
selection differ only by a known scalar constant. In particular, we
show that bias reduction in estimation is closely related to model
selection using TIC if estimation is via ML, and CLIC
\cite{varin+vidoni:2005} if estimation is via maximum composite
likelihood. Furthermore, TIC and CLIC are still consistent information
criteria when evaluated at explicit or implicit RB$M$-estimates. The
same justification we provided in Section~\ref{sec:model_selection}
for the use of information criteria at the reduced-bias estimates can
be used to justify the use of information criteria at estimates
arising from the additive adjustment of estimating functions by
alternative $O_p(1)$ quantities, like the median reduced-bias
estimates discussed in \citet{kenne+salvan+sartori:2017} and
\citet{kosmidis+kennepagui+sartori:2020}, and the mean reduced-bias
estimators in \citet{firth:1993}.

As shown in Section~\ref{sec:plugin}, the bias-reducing penalized
objectives, can be further enriched with plug-in penalties of small
asymptotic order to return RB$M$p-estimates with enhanced
properties. In Section~\ref{sec:plugin}, we used this fact to
construct reduced-bias estimates that are always finite in binomial
regression models, like logistic, probit, complementary log-log, and
cauchit regression, and in autologistic regression models for
clustered spatial data in Example~\ref{autologistic}. Note that, to
the authors' knowledge, there is no proof that the reduced-bias
estimator in \citet{firth:1993} for binomial regression models with
link functions other than logit always takes finite values; see,
\citet{kosmidis+firth:2021} for a proof of the finiteness of the
reduced-bias estimator of \citet{firth:1993} in logistic regression
when the model matrix is of full rank. 

As we discussed in Section~\ref{spatiotemporal}, when there is only
$k = 1$ observation with correlated components, application-dependent
conditions need to be used for the appropriate definition of
${\bb e}({\bb \theta})$ in the penalized objective function
in~\eqref{pen_loglik} or of ${\bb e}({\bb \theta})$ and
${\bb d}_r({\bb \theta})$ in the empirical bias-reducing
adjustment~\eqref{general_empirical_adjustment_matrix}. In the context
of time series and spatial data that do not seriously depart from the
condition of stationarity, one can consider window sub-sampling for
the definition of ${\bb e}({\bb \theta})$ and
${\bb d}_r({\bb \theta})$ (see, for example \citealt{carlstein:1986}
and \citet{heagerty+lumley:2000} for definitions and guidance on the
choice of the window size).

Our empirical experience with fully-specified models is that while the
RB$M$-estimators are first-order unbiased, they may not deliver an as
strong bias correction in small samples as other methods in
Table~\ref{br_characteristics} that require full specification of the
correct model.  Example~\ref{probit_bias} and
Section~\ref{sec:negbin_bias} in the Supplementary Material document
provide evidence about this effect.

From the simulation-based methods, the linear-bias-correcting (LBC)
estimator of \citet{mackinnon+smith:1998} has interesting properties
that may warrant further investigation. In particular, if the initial
estimator $\hat\theta$ has a linear bias function, always takes values
away from the boundary of the parameter space, and samples from the
correct model or an appropriate empirical distribution function can be
taken, then the LBC estimator of \citet{mackinnon+smith:1998} can
deliver estimators with virtually zero bias. If the assumption of
linear bias function is not satisfied, then the LBC estimator is free
from first-order bias, just like RB$M$-estimators are. These
properties of LBC come at the expense of the need to simulation from
the full model and more
computation. \citet[Section~6]{mackinnon+smith:1998} (and
Section~\ref{br_methods}) provides a clear discussion on why
estimators based on asymptotic approximation of the bias (like
RB$M$-estimators) are typically more attractive than simulation-based
estimators in applications.

The differences between RB$M$-estimators and other bias reduction
methods that aim to remove the first term or more from the bias
function in fully- or partially-specified models, are beyond the first
term of the bias expansion. Hence, while a detailed theoretical
assessment of those differences may be interesting for particular
models, it is subtle, mathematically involved and potentially
inconclusive at the level of generality RB$M$ estimation has been
developed.

\citet{lunardon:2018} showed that bias reduction in ML estimation
using the adjustments in \citet{firth:1993} can be particularly
effective for inference about a low-dimensional parameter of interest
in the presence of high-dimensional nuisance parameters, while
providing, at the same time, improved estimates of the nuisance
parameters. Current research investigates the performance of the
RB$M$-estimator for general $M$-estimation in stratified settings,
extending the optimality results in \citet{lunardon:2018} when maximum
composite likelihood or other $M$-estimators are used.

The sufficient conditions we used for RB$M$-estimation do not
necessarily cover for asymptotic regimes where
$p / n \to \kappa \in (0, 1)$ \citep[see, for
example][]{sur+candes:2019}. RB$M$-estimators have been found, though,
to deliver remarkable corrections in both bias and variance with
high-dimensional covariate specifications in partially-specified
autologistic regression (see Example~\ref{highdim_autologistic}), and
in fully-specified negative binomial regression (see
Section~\ref{sec:negbin_bias} in the Supplementary Material document). A
formal assessment of the RB$M$-estimation framework in
high-dimensional models is the subject of future work.

\section{Supplementary material}
The Supplementary Material include computer code to fully reproduce all
numerical results and figures in the paper. The code is also available
at
\url{http://www.ikosmidis.com/files/RBM\_supporting\_computer\_code.zip}.
The organization of the computer code is detailed in
Section~\ref{sec:code} of the Supplementary Material
document. Section~\ref{sec:assumptions_sup} provides the technical
conditions we employ for the theoretical developments in this work,
which are standard in the context of $M$-estimation, and an annotated
analysis of the links between them, and their wide scope and
applicability. Section~\ref{sec:supp_taylor} gives the stochastic
Taylor expansion for $\tilde{\bb \theta} - \bar{\bb \theta}$, and
Section~\ref{sec:implementation} provides a quasi Newton-Raphson
iteration for computing the RB$M$-estimates ${\bb \theta}^\dagger$ and
$\tilde{\bb \theta}$ for general models when these do not have closed
forms. Section~\ref{glm_appendix} gives the form of
${\bb j}({\bb \beta}, \phi)$ and ${\bb e}({\bb \beta}, \phi)$ for GLMs
with unknown dispersion parameter
$\phi$. Section~\ref{sec:gaussian_max} derives the mathematical
expressions that are used in Example~\ref{max_stable_ex}, and
Section~\ref{sec:additional_OLS_AR} provides results for additional
simulation experiments for $\theta = 0.2$ and $\theta = 0.9$ in the
setting of Example~\ref{OLS_AR}. Section~\ref{sec:negbin_bias}
provides results of simulation experiments that assess the performance
of RB$M$-estimation in negative binomial regression with
high-dimensional covariate specifications.

\section{Acknowledgements}
We thank David Firth and Elvezio Ronchetti for helpful discussions and
comments on an earlier version of this work. Part of this work was
completed when Nicola Lunardon was at the Department of Economics,
Management and Statistics (DEMS), University of Milano-Bicocca. The
authors acknowledge the DEMS Data Science Lab at University of
Milano-Bicocca for supporting this work by providing computational
resources. For the purpose of open access, the authors have applied a
Creative Commons Attribution (CC-BY) licence to any Author Accepted
Manuscript version arising from this submission.

\bibliographystyle{chicago}
\bibliography{robustbr}

\begin{thebibliography}{}

\bibitem[\protect\citeauthoryear{Akaike}{Akaike}{1974}]{akaike:1974}
Akaike, H. (1974).
\newblock A new look at the statistical model identification.
\newblock {\em IEEE Transactions on Automatic Control\/}~{\em 19\/}(6),
  716--723.

\bibitem[\protect\citeauthoryear{Albert and Anderson}{Albert and
  Anderson}{1984}]{albert+anderson:1984}
Albert, A. and J.~Anderson (1984).
\newblock On the existence of maximum likelihood estimates in logistic
  regression models.
\newblock {\em Biometrika\/}~{\em 71\/}(1), 1--10.

\bibitem[\protect\citeauthoryear{Bell}{Bell}{2021}]{bell:2021}
Bell, B. (2021).
\newblock {CppAD}: a package for {C}++ algorithmic differentiation (20210329).

\bibitem[\protect\citeauthoryear{Besag}{Besag}{1974}]{besag:1974}
Besag, J. (1974).
\newblock Spatial interaction and the statistical analysis of lattice systems.
\newblock {\em Journal of the Royal Statistical Society. Series B
  (Methodological)\/}~{\em 36\/}(2), 192--236.

\bibitem[\protect\citeauthoryear{Besag}{Besag}{1975}]{besag:1975}
Besag, J. (1975).
\newblock Statistical {Analysis} of {Non}-{Lattice} {Data}.
\newblock {\em Journal of the Royal Statistical Society. Series D (The
  Statistician)\/}~{\em 24\/}(3), 179--195.

\bibitem[\protect\citeauthoryear{Besag}{Besag}{1972}]{besag:1972}
Besag, J.~E. (1972).
\newblock Nearest-neighbour systems and the auto-logistic model for binary
  data.
\newblock {\em Journal of the Royal Statistical Society: Series B
  (Methodological)\/}~{\em 34\/}(1), 75--83.

\bibitem[\protect\citeauthoryear{Boos}{Boos}{1992}]{boos:1992}
Boos, D.~D. (1992).
\newblock On {Generalized} {Score} {Tests}.
\newblock {\em The American Statistician\/}~{\em 46\/}(4), 8.

\bibitem[\protect\citeauthoryear{Canty and Ripley}{Canty and
  Ripley}{2022}]{boot}
Canty, A. and B.~D. Ripley (2022).
\newblock {\em boot: Bootstrap R (S-Plus) Functions}.
\newblock R package version 1.3-28.1.

\bibitem[\protect\citeauthoryear{Caragea and Kaiser}{Caragea and
  Kaiser}{2009}]{caragea+kaiser:2009}
Caragea, P.~C. and M.~S. Kaiser (2009).
\newblock Autologistic models with interpretable parameters.
\newblock {\em Journal of Agricultural, Biological, and Environmental
  Statistics\/}~{\em 14\/}(3), 281--300.

\bibitem[\protect\citeauthoryear{Carlstein}{Carlstein}{1986}]{carlstein:1986}
Carlstein, E. (1986).
\newblock The use of subseries values for estimating the variance of a general
  statistic from a stationary sequence.
\newblock {\em The Annals of Statistics\/}~{\em 14\/}(3), 1171--1179.

\bibitem[\protect\citeauthoryear{Claeskens and Hjort}{Claeskens and
  Hjort}{2008}]{claeskens:2008}
Claeskens, G. and N.~L. Hjort (2008).
\newblock {\em Model Selection and Model Averaging}.
\newblock {Cambridge, NY}: {Cambridge University Press}.

\bibitem[\protect\citeauthoryear{Cordeiro and McCullagh}{Cordeiro and
  McCullagh}{1991}]{cordeiro+mccullagh:1991}
Cordeiro, G.~M. and P.~McCullagh (1991).
\newblock Bias correction in generalized linear models.
\newblock {\em Journal of the Royal Statistical Society. Series B
  (Methodological)\/}~{\em 53\/}(3), 629--643.

\bibitem[\protect\citeauthoryear{Davison and Gholamrezaee}{Davison and
  Gholamrezaee}{2012}]{davison+gholamrezaee:2012}
Davison, A.~C. and M.~M. Gholamrezaee (2012).
\newblock Geostatistics of extremes.
\newblock {\em Proceedings of the Royal Society A: Mathematical, Physical and
  Engineering Sciences\/}~{\em 468\/}(2138), 581--608.

\bibitem[\protect\citeauthoryear{Davison and Hinkley}{Davison and
  Hinkley}{1997}]{davison+hinkley:1997}
Davison, A.~C. and D.~V. Hinkley (1997).
\newblock {\em Bootstrap Methods and Their Application}.
\newblock {Cambridge; New York, NY}: {Cambridge University Press}.

\bibitem[\protect\citeauthoryear{Durbin}{Durbin}{1959}]{durbin:1959}
Durbin, J. (1959).
\newblock A note on the application of {Q}uenouille's method of bias reduction
  to the estimation of ratios.
\newblock {\em Biometrika\/}~{\em 46\/}(3), 477--480.

\bibitem[\protect\citeauthoryear{Efron}{Efron}{1975}]{efron:1975}
Efron, B. (1975).
\newblock Defining the curvature of a statistical problem (with applications to
  second order efficiency).
\newblock {\em The Annals of Statistics\/}~{\em 3\/}(6), 1189--1242.

\bibitem[\protect\citeauthoryear{Efron}{Efron}{1982}]{efron:1982}
Efron, B. (1982).
\newblock {\em The Jackknife, the Bootstrap and Other Resampling Plans}.
\newblock Society for Industrial and Applied Mathematics.

\bibitem[\protect\citeauthoryear{Efron and Tibshirani}{Efron and
  Tibshirani}{1993}]{efron+tibshirani:1993}
Efron, B. and R.~Tibshirani (1993).
\newblock {\em An Introduction to the Bootstrap}.
\newblock New York, NY: Chapman \& Hall/CRC.

\bibitem[\protect\citeauthoryear{Firth}{Firth}{1993}]{firth:1993}
Firth, D. (1993).
\newblock Bias reduction of maximum likelihood estimates.
\newblock {\em Biometrika\/}~{\em 80\/}(1), 27--38.

\bibitem[\protect\citeauthoryear{Genton, Ma, and Sang}{Genton
  et~al.}{2011}]{genton+ma+sang:2011}
Genton, M.~G., Y.~Ma, and H.~Sang (2011).
\newblock On the likelihood function of gaussian max-stable processes.
\newblock {\em Biometrika\/}~{\em 98\/}(2), 481--488.

\bibitem[\protect\citeauthoryear{Gourieroux, Monfort, and Renault}{Gourieroux
  et~al.}{1993}]{gourieroux:1993}
Gourieroux, C., A.~Monfort, and E.~Renault (1993).
\newblock Indirect inference.
\newblock {\em Journal of Applied Econometrics\/}~{\em 8}, S85--S118.

\bibitem[\protect\citeauthoryear{Griewank and Walther}{Griewank and
  Walther}{2008}]{griewank+walther:2008}
Griewank, A. and A.~Walther (2008).
\newblock {\em Evaluating Derivatives: Principles and Techniques of Algorithmic
  Differentiation\/} (2nd ed.).
\newblock Philadelphia, PA: Society for Industrial and Applied Mathematics.

\bibitem[\protect\citeauthoryear{Gr\"{u}n, K., and Zeileis}{Gr\"{u}n
  et~al.}{2012}]{grun+kosmidis+zeileis:2012}
Gr\"{u}n, B., I.~K., and A.~Zeileis (2012).
\newblock Extended beta regression in {R}: Shaken, stirred, mixed, and
  partitioned.
\newblock {\em Journal of Statistical Software\/}~{\em 48\/}(11), 1--25.

\bibitem[\protect\citeauthoryear{Guerrier, {Dupuis-Lozeron}, Ma, and
  {Victoria-Feser}}{Guerrier et~al.}{2019}]{guerrier+dupuis-lozeron+etal:2019}
Guerrier, S., E.~{Dupuis-Lozeron}, Y.~Ma, and M.-P. {Victoria-Feser} (2019).
\newblock Simulation-based bias correction methods for complex models.
\newblock {\em Journal of the American Statistical Association\/}~{\em
  114\/}(525), 146--157.

\bibitem[\protect\citeauthoryear{Hall and Martin}{Hall and
  Martin}{1988}]{hall+martin:1988}
Hall, P. and M.~A. Martin (1988).
\newblock On bootstrap resampling and iteration.
\newblock {\em Biometrika\/}~{\em 75\/}(4), 661--671.

\bibitem[\protect\citeauthoryear{Heagerty and Lumley}{Heagerty and
  Lumley}{2000}]{heagerty+lumley:2000}
Heagerty, P.~J. and T.~Lumley (2000).
\newblock Window subsampling of estimating functions with application to
  regression models.
\newblock {\em Journal of the American Statistical Association\/}~{\em
  95\/}(449), 197--211.

\bibitem[\protect\citeauthoryear{Hughes, Haran, and Caragea}{Hughes
  et~al.}{2011}]{hughes+haran+caragea:2011}
Hughes, J., M.~Haran, and P.~C. Caragea (2011).
\newblock Autologistic models for binary data on a lattice.
\newblock {\em Environmetrics\/}~{\em 22\/}(7), 857--871.

\bibitem[\protect\citeauthoryear{Huser and Davison}{Huser and
  Davison}{2013}]{huser+davison:2013}
Huser, R. and A.~C. Davison (2013).
\newblock Composite likelihood estimation for the {B}rown--{R}esnick process.
\newblock {\em Biometrika\/}~{\em 100\/}(2), 511--518.

\bibitem[\protect\citeauthoryear{Kenne~Pagui, Salvan, and Sartori}{Kenne~Pagui
  et~al.}{2017}]{kenne+salvan+sartori:2017}
Kenne~Pagui, E.~C., A.~Salvan, and N.~Sartori (2017).
\newblock Median bias reduction of maximum likelihood estimates.
\newblock {\em Biometrika\/}~{\em 104\/}(4), 923--938.

\bibitem[\protect\citeauthoryear{Konis}{Konis}{2007}]{konis:2007}
Konis, K. (2007).
\newblock {\em Linear programming algorithms for detecting separated data in
  binary logistic regression models}.
\newblock Ph.\ D. thesis, University of Oxford.

\bibitem[\protect\citeauthoryear{Kosmidis}{Kosmidis}{2014}]{kosmidis:2014}
Kosmidis, I. (2014).
\newblock Bias in parametric estimation: reduction and useful side-effects.
\newblock {\em Wiley Interdisciplinary Reviews: Computational
  Statistics\/}~{\em 6\/}(3), 185--196.

\bibitem[\protect\citeauthoryear{Kosmidis}{Kosmidis}{2023}]{brglm2}
Kosmidis, I. (2023).
\newblock {\em {brglm2}: Bias Reduction in Generalized Linear Models}.
\newblock R package version 0.9.

\bibitem[\protect\citeauthoryear{Kosmidis and Firth}{Kosmidis and
  Firth}{2009}]{kosmidis+firth:2009}
Kosmidis, I. and D.~Firth (2009).
\newblock Bias reduction in exponential family nonlinear models.
\newblock {\em Biometrika\/}~{\em 96\/}(4), 793--804.

\bibitem[\protect\citeauthoryear{Kosmidis and Firth}{Kosmidis and
  Firth}{2021}]{kosmidis+firth:2021}
Kosmidis, I. and D.~Firth (2021).
\newblock Jeffreys-prior penalty, finiteness and shrinkage in binomial-response
  generalized linear models.
\newblock {\em Biometrika\/}~{\em 108\/}(1), 71--82.

\bibitem[\protect\citeauthoryear{Kosmidis, Kenne~Pagui, and Sartori}{Kosmidis
  et~al.}{2020}]{kosmidis+kennepagui+sartori:2020}
Kosmidis, I., E.~C. Kenne~Pagui, and N.~Sartori (2020).
\newblock Mean and median bias reduction in generalized linear models.
\newblock {\em Statistics and Computing\/}~{\em 30\/}(1), 43--59.

\bibitem[\protect\citeauthoryear{Kosmidis and Lunardon}{Kosmidis and
  Lunardon}{2022}]{mestimation}
Kosmidis, I. and N.~Lunardon (2022).
\newblock {\em {MEstimation.jl}: Methods for {M}-estimation of statistical
  models}.
\newblock Julia package version 0.2.0.

\bibitem[\protect\citeauthoryear{Kosmidis, Schumacher, and
  Schwendinger}{Kosmidis et~al.}{2022}]{detectseparation}
Kosmidis, I., D.~Schumacher, and F.~Schwendinger (2022).
\newblock {\em detectseparation: Detect and Check for Separation and Infinite
  Maximum Likelihood Estimates}.
\newblock R package version 0.3.

\bibitem[\protect\citeauthoryear{Kristensen, Nielsen, Berg, Skaug, and
  Bell}{Kristensen et~al.}{2016}]{kristensen+nielsen+berg+etal:2016}
Kristensen, K., A.~Nielsen, C.~W. Berg, H.~Skaug, and B.~M. Bell (2016).
\newblock {TMB}: Automatic differentiation and {L}aplace approximation.
\newblock {\em Journal of Statistical Software\/}~{\em 70\/}(5), 1--21.

\bibitem[\protect\citeauthoryear{Kuk}{Kuk}{1995}]{kuk:1995}
Kuk, A. Y.~C. (1995).
\newblock Asymptotically unbiased estimation in generalized linear models with
  random effects.
\newblock {\em Journal of the Royal Statistical Society. Series B
  (Methodological)\/}~{\em 57}, 395--407.

\bibitem[\protect\citeauthoryear{Kunsch}{Kunsch}{1989}]{kunsch:1989}
Kunsch, H.~R. (1989).
\newblock The jackknife and the bootstrap for general stationary observations.
\newblock {\em The Annals of Statistics\/}~{\em 17\/}(3), 1217--1241.

\bibitem[\protect\citeauthoryear{Liang and Zeger}{Liang and
  Zeger}{1986}]{liang+zeger:1986}
Liang, K.-Y. and S.~L. Zeger (1986).
\newblock Longitudinal data analysis using generalized linear models.
\newblock {\em Biometrika\/}~{\em 73\/}(1), 13--22.

\bibitem[\protect\citeauthoryear{Lindsay}{Lindsay}{1988}]{lindsay:1988}
Lindsay, B.~G. (1988).
\newblock Composite likelihood methods.
\newblock In N.~U. Prabhu (Ed.), {\em Statistical {Inference} from {Stochastic}
  {Processes}}, Volume~80 of {\em Contemporary {Mathematics}}. American
  Mathematical Society.

\bibitem[\protect\citeauthoryear{Lunardon}{Lunardon}{2018}]{lunardon:2018}
Lunardon, N. (2018).
\newblock On bias reduction and incidental parameters.
\newblock {\em Biometrika\/}~{\em 105\/}(1), 233--238.

\bibitem[\protect\citeauthoryear{Lunardon and Scharfstein}{Lunardon and
  Scharfstein}{2017}]{lunardon+scharfstein:2017}
Lunardon, N. and D.~Scharfstein (2017).
\newblock Comment on ``{S}mall sample {GEE} estimation of regression parameters
  for longitudinal data".
\newblock {\em Statistics in medicine\/}~{\em 36\/}(22), 3596--3600.

\bibitem[\protect\citeauthoryear{MacKinnon and Smith}{MacKinnon and
  Smith}{1998}]{mackinnon+smith:1998}
MacKinnon, J.~G. and A.~A. Smith (1998).
\newblock Approximate bias correction in econometrics.
\newblock {\em Journal of Econometrics\/}~{\em 85\/}(2), 205--230.

\bibitem[\protect\citeauthoryear{Mansournia, Geroldinger, Greenland, and
  Heinze}{Mansournia et~al.}{2018}]{mansournia+etal:2018}
Mansournia, M.~A., A.~Geroldinger, S.~Greenland, and G.~Heinze (2018).
\newblock Separation in {Logistic} {Regression}: {Causes}, {Consequences}, and
  {Control}.
\newblock {\em American Journal of Epidemiology\/}~{\em 187\/}(4), 864--870.

\bibitem[\protect\citeauthoryear{McCullagh}{McCullagh}{2018}]{mccullagh:2018}
McCullagh, P. (2018).
\newblock {\em Tensor methods in statistics\/} (2nd ed.).
\newblock Mineola, NY: Dover Publications.

\bibitem[\protect\citeauthoryear{Mogensen and Riseth}{Mogensen and
  Riseth}{2018}]{optim}
Mogensen, P.~K. and A.~N. Riseth (2018).
\newblock Optim: A mathematical optimization package for {Julia}.
\newblock {\em Journal of Open Source Software\/}~{\em 3\/}(24), 615.

\bibitem[\protect\citeauthoryear{Newey and Smith}{Newey and
  Smith}{2004}]{newey+smith:2004}
Newey, W.~K. and R.~J. Smith (2004).
\newblock Higher order properties of {GMM} and generalized empirical likelihood
  estimators.
\newblock {\em Econometrica\/}~{\em 72\/}(1), 219--255.

\bibitem[\protect\citeauthoryear{Pace and Salvan}{Pace and
  Salvan}{1997}]{pace+salvan:1997}
Pace, L. and A.~Salvan (1997).
\newblock {\em Principles of Statistical Inference from a Neo-{F}isherian
  Perspective}.
\newblock London, UK: {W}orld {S}cientific.

\bibitem[\protect\citeauthoryear{Padoan and Bevilacqua}{Padoan and
  Bevilacqua}{2015}]{padoan+bevilacqua:2015}
Padoan, S.~A. and M.~Bevilacqua (2015).
\newblock Analysis of random fields using {CompRandFld}.
\newblock {\em Journal of Statistical Software\/}~{\em 63\/}(9), 1--27.

\bibitem[\protect\citeauthoryear{Padoan, Ribatet, and Sisson}{Padoan
  et~al.}{2010}]{padoan+ribadet+sisson:2010}
Padoan, S.~A., M.~Ribatet, and S.~A. Sisson (2010).
\newblock Likelihood-based inference for max-stable processes.
\newblock {\em Journal of the American Statistical Association\/}~{\em
  105\/}(489), 263--277.

\bibitem[\protect\citeauthoryear{Politis and Romano}{Politis and
  Romano}{1994}]{politis+romano:1994}
Politis, D.~N. and J.~P. Romano (1994).
\newblock The stationary bootstrap.
\newblock {\em Journal of the American Statistical association\/}~{\em
  89\/}(428), 1303--1313.

\bibitem[\protect\citeauthoryear{Pratt}{Pratt}{1981}]{pratt:1981}
Pratt, J.~W. (1981).
\newblock Concavity of the log likelihood.
\newblock {\em Journal of the American Statistical Association\/}~{\em
  76\/}(373), 103--106.

\bibitem[\protect\citeauthoryear{Quenouille}{Quenouille}{1949}]{quenouille:1949}
Quenouille, M.~H. (1949).
\newblock Approximate tests of correlation in time-series 3.
\newblock {\em Mathematical Proceedings of the Cambridge Philosophical
  Society\/}~{\em 45\/}(3), 483–484.

\bibitem[\protect\citeauthoryear{Quenouille}{Quenouille}{1956}]{quenouille:1956}
Quenouille, M.~H. (1956).
\newblock Notes on bias in estimation.
\newblock {\em Biometrika\/}~{\em 43\/}(3), 353--360.

\bibitem[\protect\citeauthoryear{{R Core Team}}{{R Core Team}}{2021}]{rproject}
{R Core Team} (2021).
\newblock {\em R: A Language and Environment for Statistical Computing}.
\newblock Vienna, Austria: R Foundation for Statistical Computing.

\bibitem[\protect\citeauthoryear{Revels, Lubin, and Papamarkou}{Revels
  et~al.}{2016}]{revels+lubin+papamarkou:2016}
Revels, J., M.~Lubin, and T.~Papamarkou (2016).
\newblock Forward-mode automatic differentiation in julia.
\newblock {\em arXiv e-prints\/}, arXiv:1607.07892.

\bibitem[\protect\citeauthoryear{{Ribeiro Jr}, Diggle, Christensen, Schlather,
  Bivand, and Ripley}{{Ribeiro Jr} et~al.}{2022}]{geoR}
{Ribeiro Jr}, P.~J., P.~Diggle, O.~Christensen, M.~Schlather, R.~Bivand, and
  B.~Ripley (2022).
\newblock {\em geoR: Analysis of Geostatistical Data}.
\newblock R package version 1.9-2.

\bibitem[\protect\citeauthoryear{Stefanski and Boos}{Stefanski and
  Boos}{2002}]{stefanski+boos:2002}
Stefanski, L.~A. and D.~D. Boos (2002).
\newblock The calculus of {M}-estimation.
\newblock {\em The American Statistician\/}~{\em 56\/}(1), 29--38.

\bibitem[\protect\citeauthoryear{Sur and Cand{\`e}s}{Sur and
  Cand{\`e}s}{2019}]{sur+candes:2019}
Sur, P. and E.~J. Cand{\`e}s (2019).
\newblock A modern maximum-likelihood theory for high-dimensional logistic
  regression.
\newblock {\em Proceedings of the National Academy of Sciences\/}~{\em
  116\/}(29), 14516--14525.

\bibitem[\protect\citeauthoryear{Takeuchi}{Takeuchi}{1976}]{takeuchi:1976}
Takeuchi, K. (1976).
\newblock Distribution of information statistics and validity criteria of
  models.
\newblock {\em Mathematical Science\/}~{\em 153}, 12--18.

\bibitem[\protect\citeauthoryear{Thomson, Connor, D'Alessandro, Rowlingson,
  Diggle, Cresswell, and Greenwood}{Thomson et~al.}{1999}]{thompson+etal:1999}
Thomson, M.~C., S.~J. Connor, U.~D'Alessandro, B.~Rowlingson, P.~Diggle,
  M.~Cresswell, and B.~Greenwood (1999).
\newblock Predicting malaria infection in {Gambian} children from satellite
  data and bed net use surveys: the importance of spatial correlation in the
  interpretation of results.
\newblock {\em The American Journal of Tropical Medicine and Hygiene\/}~{\em
  61\/}(1), 2--8.

\bibitem[\protect\citeauthoryear{van~der Vaart}{van~der
  Vaart}{1998}]{vaart:1998}
van~der Vaart, A.~W. (1998).
\newblock {\em Asymptotic Statistics}.
\newblock Cambridge, UK: Cambridge University Press.

\bibitem[\protect\citeauthoryear{Varin, Reid, and Firth}{Varin
  et~al.}{2011}]{varin+reid+firth:2011}
Varin, C., N.~Reid, and D.~Firth (2011).
\newblock An overview of composite likelihood methods.
\newblock {\em Statistica Sinica\/}~{\em 21\/}(1), 5--42.

\bibitem[\protect\citeauthoryear{Varin and Vidoni}{Varin and
  Vidoni}{2005}]{varin+vidoni:2005}
Varin, C. and P.~Vidoni (2005).
\newblock A note on composite likelihood inference and model selection.
\newblock {\em Biometrika\/}~{\em 92\/}(3), 519--528.

\bibitem[\protect\citeauthoryear{Wedderburn}{Wedderburn}{1974}]{wedderburn:1974}
Wedderburn, R. (1974).
\newblock Quasi-likelihood functions, generalized linear models, and the
  {G}auss-{N}ewton method.
\newblock {\em Biometrika\/}~{\em 61\/}(3), 439--447.

\bibitem[\protect\citeauthoryear{Wolters}{Wolters}{2022}]{autologistic}
Wolters, M. (2022).
\newblock {\em {Autologistic.jl}}.
\newblock Julia package version 0.5.1.

\bibitem[\protect\citeauthoryear{Wolters}{Wolters}{2017}]{wolters:2017}
Wolters, M.~A. (2017).
\newblock Better autologistic regression.
\newblock {\em Frontiers in Applied Mathematics and Statistics\/}~{\em 3}, 24.

\end{thebibliography}

\includepdf[pages=-]{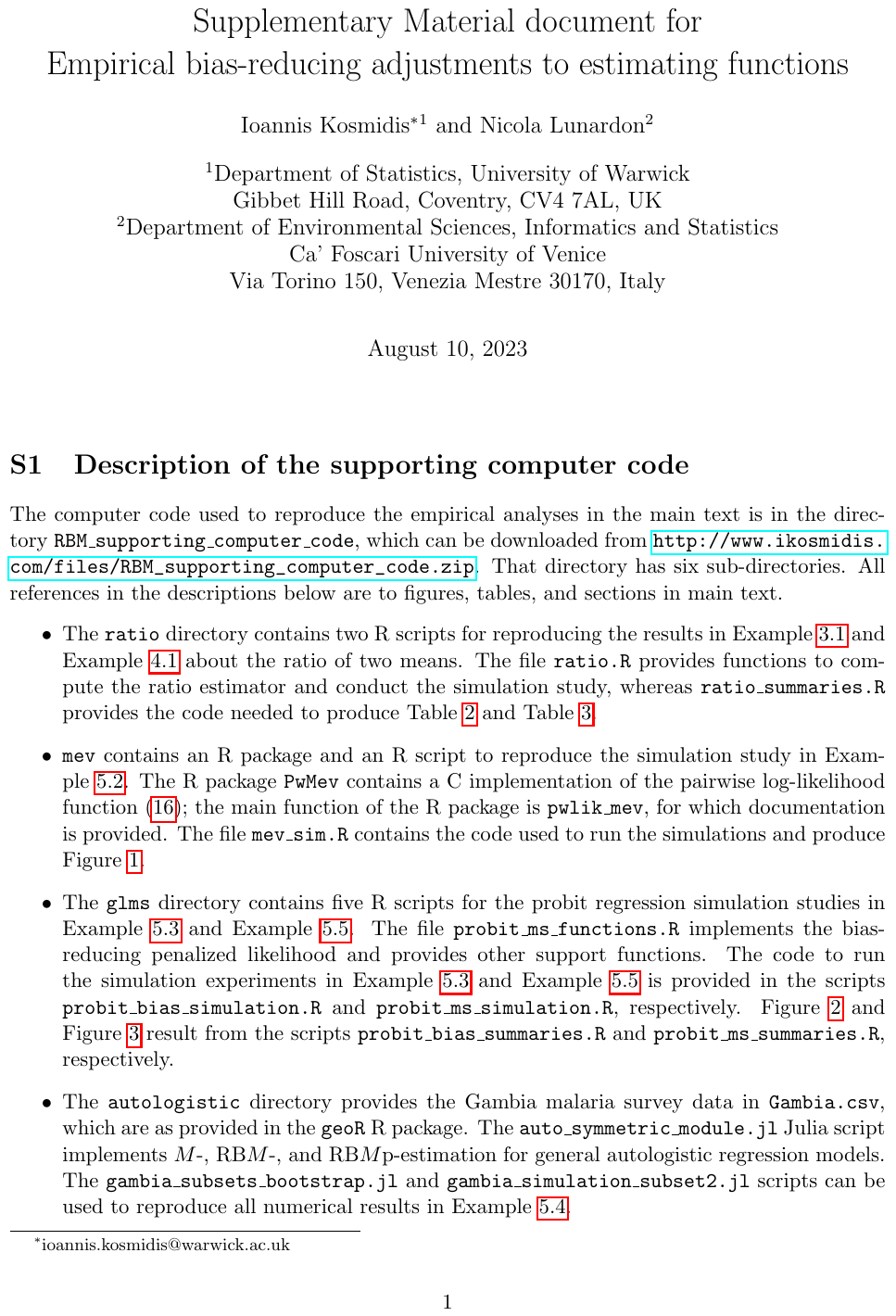}

\end{document}